\documentclass[10pt]{article}
\usepackage[a4paper,margin=1in]{geometry}
\usepackage{amsmath,amssymb,amsthm,bm} 
\usepackage{mathtools}
\usepackage{enumitem}
\usepackage{booktabs}
\usepackage{hyperref}
\usepackage{xcolor}
\usepackage{authblk}
\usepackage{longtable}
\usepackage[square,numbers]{natbib}
\usepackage{pdfpages}

\title{Estimands and estimation in trials with time trends}
\author[1, $*$]{Marta Bofill Roig}
\author[2]{Ekkehard Glimm}
\author[3, 4]{Kelly Van Lancker}
\author[5]{Martin Posch}

\affil[1]{{Department of Statistics and Operations Research and Institute for Research and Innovation in Health (IRIS)}, {Universitat Politècnica de Catalunya - BarcelonaTech}, {{Barcelona}, {Spain}}}

\affil[2]{{Advanced Methodology and Data Science}, {Novartis Pharma AG}, {{Basel}, {Switzerland}}}

\affil[3]{{Department of Mathematics, Computer Science and Statistics}, {Ghent University}, {{Ghent}, {Belgium}}}

\affil[4]{{Department of Mathematics and Data Science}, {Vrije Universiteit Brussel}, {{Brussels}, {Belgium}}}

\affil[5]{{Center for Medical Data Science}, {Medical University of Vienna}, {{Vienna}, {Austria}}}

\affil[$*$]{marta.bofill.roig@upc.edu}
\date{}

\begin{document}

\maketitle 

\begin{abstract}

In platform trials, model-based approaches typically estimate treatment effects conditional on calendar time. However, scientific and regulatory interest often lies in treatment effects defined for a target population spanning multiple enrollment periods, requiring explicit consideration of how effects should be averaged across time. 
This raises two fundamental challenges. The first is the definition of the appropriate target estimand when combining data across multiple periods. The second is the selection of the estimator to be used. In this work, we examine conditional and marginal estimands in platform trials with time trends, and describe target populations of interest. To address the second challenge, we evaluate model-based, G-computation and augmented inverse probability weighting estimators, comparing their bias and variance. We discuss how the choice of estimand, target population and trial data used for estimation affects estimator performance.

\end{abstract}

\section{Introduction}\label{sec-intro} 

Platform trials are multi-arm, multi-stage trials in which treatment arms may enter or leave the trial over time while sharing a common control group\cite{berry2015platform,saville2016efficiencies,koenig2024current}. Their flexibility can substantially improve efficiency compared with conventional trials by enabling the evaluation of multiple treatments within a single infrastructure. Because platform trials often operate over long periods of time, changes in patient characteristics, standards of care, disease prevalence, or other temporal factors may induce time trends. If not properly accounted for, such trends can bias treatment effect estimates. Consequently, analyses often require adjustment for time, particularly when non-concurrent controls are incorporated.

A major methodological development in recent years has therefore focused on the use of period-adjusted analyses. The potential advantages and risks of using non-concurrent controls in the analysis are now well understood \cite{sridhara2022use,burgwinkel2023use}. On the one hand, they may improve precision of treatment effect estimates by increasing the amount of available control information. On the other hand, naive comparisons may be biased when outcomes or patient characteristics evolve over calendar time. A considerable methodological literature has therefore focused on methods that account for temporal trends. Model-based approaches that incorporate time (period) as a covariate are generally the preferred solution because, under assumptions such as the absence of treatment-by-period interactions,  they can provide (asymptotically) unbiased treatment effect estimators despite the presence of time trends \cite{lee2020including,bofill2022model, saville2022bayesian, marschner2024analysis, guo2024treatment,zhan2025integrated, krotka2025statistical}. Alternative data-borrowing methods, similar to those used in historical-control settings, have also been proposed, although these generally do not guarantee unbiased estimation in the presence of temporal drift \cite{kopp2020power}. Importantly, adjustment for period is not only relevant when non-concurrent controls are used. Even analyses restricted to concurrent controls may require period adjustment when allocation ratios change over the course of the study, as such changes can themselves induce bias if temporal trends are present. 

While much of the methodological literature has focused on how to obtain unbiased treatment effect estimates in the presence of time trends, a more fundamental question has received comparatively less attention: what treatment effect should be estimated? In randomized trials, recent work on covariate adjustment has highlighted the importance of clearly specifying the estimand before selecting an estimation strategy \cite{tsiatis2008covariate, van2024covariate,bannick2026covariate}. In particular, a distinction must be made between conditional treatment effects, which are defined within strata of covariates or time periods, and marginal treatment effects, which average treatment effects over a target population \cite{fda2023covariate}.    
The choice between these estimands is particularly important in platform trials because treatment availability changes over calendar time. 

Indeed, a defining feature of platform trials is that different treatment comparisons may involve different subsets of patients. For a treatment arm that enters or exits the platform, only a subset of enrolled patients could ever have been randomized to that treatment. Consequently, multiple target populations can be defined. One may focus on all patients who were concurrently eligible for a treatment comparison, only those actually assigned to the treatment or control under investigation, or only patients enrolled during the most recent period in which the treatment was available. These populations need not coincide,  
and can lead to different treatment effects and therefore different estimands.

Recent methodological work has begun to address these questions. 
Recognizing that the staggered entry and exit of treatment arms fundamentally modifies the population underlying a treatment comparison, Qian et al. \cite{qian2024estimands} considered both the definition of an appropriate estimand and its estimation. They proposed the \textit{Entire Concurrently Eligible} (ECE) population, consisting of all individuals who were eligible for both the experimental treatment under evaluation and the shared control, and who entered the trial during periods in which both were available for randomization. Thus, the ECE population includes not only participants randomized to the treatment of interest or control, but also participants assigned to other active treatment arms, provided they were eligible for the comparison of interest. The authors argued that this is the right baseline population because it preserves the randomized comparison, respects eligibility criteria, and is invariant to operational trial characteristics such as randomization ratios. In addition, they presented estimators together with corresponding asymptotic results. They focus on the concurrent population and do not incorporate non-concurrent control data in the analysis.  

Santacatterina et al. \cite{santacatterina2025identification} approached the problem from an identification perspective. Rather than starting from estimation procedures, they first discussed the assumptions required to identify treatment effects when non-concurrent controls are incorporated. 
Their preferred estimand is a concurrent average treatment effect, namely an intention-to-treat marginal causal effect among participants enrolled while both the treatment under evaluation and the shared control were available for assignment. 
Although motivated differently, this estimand is closely related to the ECE framework in that it focuses on treatment effects within concurrent populations.

Once an estimand has been defined, a second question concerns estimation. Several approaches have been proposed, including regression adjustment, standardization (G-computation), inverse probability weighting, and augmented inverse probability weighting (AIPW) \cite{fda2023covariate,qian2024estimands,santacatterina2025identification}. These methods differ not only in their robustness and efficiency properties but also in the amount of trial data they use. Importantly, the target population defining the estimand must be distinguished from the analysis dataset used for estimation. 

Despite these important advances, the relationship between estimands, target populations, and estimation strategies in platform trials remains incompletely understood. While recent work has clarified important aspects of estimand definition and identification, several fundamental questions remain unresolved.
First, it is unclear under which conditions different concurrent target populations correspond to the same treatment effect and under which conditions they represent genuinely different estimands.
Second, when these estimands differ, the impact of the chosen target population on statistical precision and inference remains poorly understood. Third, the distinction between the target population defining the estimand and the analysis dataset used for estimation is often blurred, making it difficult to disentangle whether gains in efficiency arise from targeting a different population or simply from using additional data for estimation. Finally, while these issues are relatively straightforward for continuous outcomes under linear models, they become considerably more complex for non-linear effect measures, where marginal treatment effects may vary across periods even when conditional treatment effects remain constant.

In this work, we investigate both the definition (Section~3) and estimation (Section~4) of treatment effects in platform trials with time trends. We first characterize several target populations of scientific interest and the corresponding conditional and marginal estimands. We establish conditions under which commonly considered concurrent estimands coincide and identify situations in which they differ. We then study the statistical properties of the unadjusted estimators, regression-based estimators, standardization (G-computation), and augmented inverse probability weighting estimators. 
Beyond the definition of estimands and estimators, we derive analytical results that clarify the relationship between target populations, analysis datasets, and statistical efficiency (Section~5). In particular, we investigate the consistency of the estimators, establish conditions under which different target populations lead to identical estimands, derive variance expressions and efficiency comparisons, and examine how the use of concurrent and non-concurrent data affects precision. For adjusted estimators, we explicitly distinguish the role of the target population from that of the analysis dataset and show how each contributes to estimator variance. 
Finally, through simulation studies and numerical examples for both continuous and binary outcomes (Sections~6 and~7), we evaluate how the choice of estimand, target population, estimator, and analysis dataset influences bias, precision, and variance estimation.

\section{Trial framework and notation} \label{sec:notation}

\begin{figure}[h!]
	\centering
	\includegraphics[width=0.8\linewidth,page=3]{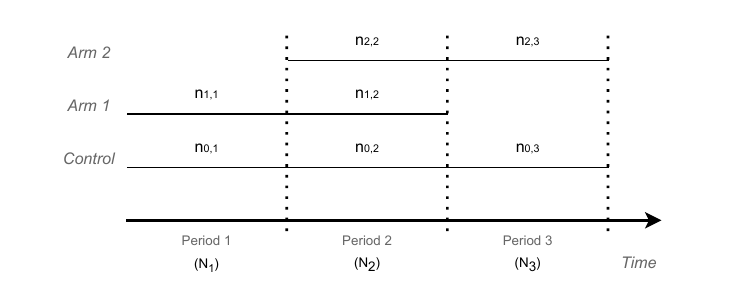}
	\caption{Graphical representation of the trial design}
	\label{fig:example}
\end{figure}

For educational reasons, we consider the following simple setup: a platform trial with two experimental treatments ($a=1,2$) and a shared control ($a=0$). The second experimental arm ($a=2$) becomes available only at a later stage, while the first experimental arm ($a=1$) is discontinued before the trial ends.  
We divide trial duration into periods, defined as time intervals during which the set of available treatment arms remains unchanged together with the allocation ratios. Period 1 begins at trial initiation with the control and arm 1 available. Period 2 begins when arm 2 is added. After two periods, arm 1 is discontinued and is therefore no longer available in period 3. See Figure \ref{fig:example} for an illustration of the design. 

For a focal comparison of treatment $a$ versus control, let $C_a$ denote the index set of periods where arm $a$ is active (i.e., available for randomization),  $L_a$ the last period where arm $a$ is active, and $U_a$ all periods from trial start through the last period arm $a$ is active. For example, for the comparison of treatment 2 versus control  $C_2=\{2,3\}$,  $L_2=3$ and $U_2=\{1,2,3\}$.

Let $Y$ be the outcome, $S$ the calendar period, and $A$ the treatment assignment. Let $X$ denote a vector of baseline covariates. Let $Y^{(a)}$ denote the potential outcome under treatment $a$. 
%
Let $n_{a,s}$ denote the number randomized to treatment $a$ in period $s$, and $n_{0,s}$ the number randomized to control in period $s$.
Define $m_{a,s} = n_{a,s} + n_{0,s}$ as the number assigned to arm $a$ or control in period $s$.  
Within each period $s$, the covariates and potential outcomes $(X_i, Y_i^{(0)}, Y_i^{(1)}, Y_i^{(2)})$ of the $N_s$ participants enrolled in that period are independent draws from a period-specific distribution. The period sizes $N_s$ are treated as fixed and, because block randomization is used within periods, so are $n_{a,s}$ and $m_{a,s} = n_{a,s} + n_{0,s}$.  

We make the following assumptions. (i) consistency, i.e. $Y_i = Y_i^{(A_i)}$ for all participants; (ii) positivity in concurrent periods, $P(A = a \mid S = s) > 0$ and $P(A = 0 \mid S = s) > 0$ for all $s \in C_a$; 
and finally, (iii) randomization within period, meaning that $A$ is assigned such that $A$ is stochastically independent of potential outcomes and $X$ conditional on $S$, that is, $A \perp \{Y^{(0)}, Y^{(1)}, Y^{(2)}, X\} \mid S$. 

For a continuous outcome, define
$$\Delta_{a,s}(x) = 
E[Y^a-Y^0 | S=s, X=x] 
$$
as the conditional period-specific treatment effect for arm $a$ in period $s$, given covariates $X=x$. 

The corresponding
marginal period-specific treatment effect is obtained by averaging over the covariate distribution:
$$\Delta_{a,s} = E_{X}[\Delta_{a,s}(X)|S=s] $$
The binary outcome case is treated separately in Section~\ref{sec:binary}. 

\section{Estimands and target populations} \label{sec:estimands}

In platform trials, several populations are of interest due to staggered entry and exit of treatment arms. We distinguish the following populations:
\begin{enumerate} 
	\item \textbf{Entire concurrently eligible (ECE) population}: Patients with a positive probability of being assigned to treatment $a$ at their time of entry, regardless of the treatment to which they were actually assigned. In Figure \ref{fig:example}, for treatment $a=1$ versus $a=0$, the patients corresponding to $n_{0,1}+n_{1,1}+n_{0,2}+n_{1,2}+n_{2,2}$ constitute a random sample from this population;
	\item \textbf{Actually concurrently assigned (ACA) population}: Patients with a positive probability of being assigned to treatment $a$ who were actually assigned to either treatment $a$ or control. In Figure \ref{fig:example}, for treatment $a=1$ versus $a=0$, the patients corresponding to $n_{0,1}+n_{1,1}+n_{0,2}+n_{1,2}$ constitute a sample from this population; 
	\item \textbf{Latest concurrently eligible (LCE) population}: Patients with a positive probability of being assigned to treatment $a$, regardless of the treatment to which they were actually assigned, during the last period in which arm $a$ was available. In Figure \ref{fig:example}, for treatment $a=1$ versus $a=0$, the patients corresponding to $n_{0,2}+n_{1,2}+n_{2,2}$ constitute a random sample from this population.
	\item \textbf{Latest actually concurrently assigned (LACA) population}: Patients with a positive probability of being assigned to treatment $a$ who were actually assigned to either treatment $a$ or control during the last period in which  arm $a$ was available. In Figure \ref{fig:example}, for treatment $a=1$ versus $a=0$, the patients corresponding to $n_{0,2}+n_{1,2}$ constitute a random sample from this population.
	\item \textbf{Entire eligible (EP) population}: All patients from trial start through the last period arm $a$ is active, regardless of the treatment or period regardless of the treatment they received or their enrolment period. In Figure \ref{fig:example}, for treatment $a=2$ versus $a=0$, the patients corresponding to $N_1+N_2+N_3$ constitute a random sample from this population. Also note that for treatment $a=1$ versus $a=0$,  ECE and EP populations coincide.
\end{enumerate} 
Note that all populations are time-dependent and that, except for EP, all populations focus on periods with concurrent treatments only. 
The ECE population is the broadest among the concurrent populations, as it includes patients assigned to other experimental treatments in addition to $a$ and control. This target population corresponds to the ``Entire Concurrently Eligible Population'' defined by Qian et al. \cite{qian2024estimands}. 


Also note that the target population is identical for both LCE and LACA due to randomisation. 
While the target population remains constant, the choice between LCE and LACA as the analysis sample can lead to different statistical properties, particularly regarding estimator efficiency. The distinction lies in the sets of patients used for the analysis. LCE uses a broader sample, retaining all eligible patients from that period, thus, including those randomised to other experimental arms (e.g., arm 2 in LCE when considering the comparison treatment $a=1$ versus $a=0$). In this work, we mainly focus on concurrent population and consider LACA over LCE. 

\subsection{Concurrent estimands}

Denoting by $(P)$ the population of interest, we can write all marginal concurrent estimands as:
\begin{equation}\label{estimandpop}
	\Delta_a^{(P)} = \sum_{s \in C_a} w_{a,s}^{(P)} \Delta_{a,s},
	\qquad
	\sum_{s \in C_a} w_{a,s}^{(P)} = 1.
\end{equation}
where the weights $w_{a,s}^{(P)}$ differ across populations $(P)$,  yielding  different estimands $\Delta_a^{(P)}$. 
Throughout, $\Delta_{a,s}$ denotes this covariate-marginal effect within period $s$, and thus $\Delta_a^{(P)} = E_{(P)}\{\Delta_{a,S}(X)\}$, where the
expectation is over the joint distribution of $(S,X)$ in the target population. 
Conditional effects $\Delta_{a,s}(x)$ appear only in the binary case (Section~\ref{sec:binary}).

The weights $w_{a,s}^{(P)}$ are defined as follows:
\begin{enumerate}
	\item ECE estimand: Weights are proportional to the number of patients eligible for arm $a$ in each period:
	\[
	w_{a,s}^{\mathrm{ECE}} = \frac{N_s}{\sum_{t \in C_a} N_t},
	\qquad s \in C_a.
	\]
	\item ACA estimand: Weights are proportional to the number of patients actually randomized to either arm $a$ or control:
	\[
	w_{a,s}^{\mathrm{ACA}} = \frac{m_{a,s}}{\sum_{t \in C_a} m_{a,t}},
	\qquad s \in C_a.
	\]
	\item LACA estimand: This estimand focuses on the last period in which arm $a$ is active, then $$w_{a,L_a}^{\mathrm{LACA}}=1$$ 
	and $w_{a,s}^{\mathrm{LACA}}=0$ for $s\neq L_a$. In other words, only the last period contributes. Hence, $\Delta_a^{\mathrm{LACA}} = \Delta_{a,L_a}.$ 
\end{enumerate}

\subsection{Entire eligible population (EP) estimand}

Define a target population consisting of all patients enrolled in the platform up to the time arm $a$ leaves the platform. 
The estimand is
\[
\Delta_a^{\mathrm{EP}} = \sum_{s \in U_a} w_{a,s}^{\mathrm{EP}} \Delta_{a,s},
\qquad
\sum_{s \in U_a} w_{a,s}^{\mathrm{EP}} = 1,
\]
with weights proportional to the total platform sample size in each period:
\[
w_{a,s}^{\mathrm{EP}} = \frac{N_s}{\sum_{t \in U_a} N_t},
\qquad s \in U_a.
\]
This estimand requires an additional remark. If arm $a^*$ was not available in period $s^*\in U_a$, a model which extrapolates $\Delta_{a^*,s^*}$ from the $\Delta_{a,s}$ of other periods or other treatments is needed. The easiest such model arises when the treatment effect is assumed to be constant across all periods in $U_a$, i.e.
\[
\Delta_{a,s} = \Delta_{a,s'} \qquad \text{for all } s,s' \in U_a,
\]
which corresponds to the absence of treatment-by-period interaction, together with the standard assumptions used for non-concurrent borrowing. 

As highlighted by \cite{santacatterina2025identification}, targeting an average treatment effect over the entire duration of the trial-- including non-concurrent periods-- is potentially ``dangerous'' because it relies on untestable extrapolation. Specifically, while the exchangeability of the control outcome mechanism across periods (Assumption 5 in \cite{santacatterina2025identification}) can often be empirically checked or satisfied via correct model specification, the extrapolation of the treated outcome mechanism to periods where the arm was not available (Assumption 6 in \cite{santacatterina2025identification}) cannot be verified from the observed data. Consequently, the validity of $\Delta_a^{\mathrm{EP}}$ depends on the assumption that the expected outcome under treatment $a$, conditional on baseline covariates, can be reliably extrapolated from the concurrent periods to the non-concurrent periods in $U_a$.

While we agree that the focus should remain on estimands defined over concurrent periods to avoid these untestable assumptions, the entire eligible population (including non-concurrent controls) can still be utilized as an analysis set. Specifically, fitting outcome models to this broader dataset can improve the estimation of the relationship between covariates and outcomes for concurrent estimands, thereby increasing statistical efficiency in finite samples. In Sections \ref{sec:rescont} and \ref{sec:simcont}, we maintain this distinction by focusing on concurrent estimands while considering the efficiency gains offered by larger analysis samples. 

\section{Point estimators for treatment effects}\label{sec:estimators}

In this section, we describe estimators for the treatment effects defined in Section~\ref{sec:estimands}. Specifically, we consider an unadjusted estimator, a regression-based estimator, a G-computation estimator and an augmented inverse probability weighting estimator.

Recall from Section~\ref{sec:estimands} that every population-level estimand takes the form $\Delta_a^{(P)} = \sum_{s} w_{a,s}^{(P)} \Delta_{a,s}$, with weights $w_{a,s}^{(P)}$ depending on the target population $(P)$. The unadjusted, G-computation and augmented inverse probability weighting estimators can be expressed as
\begin{equation}\label{eq:generic}
	\hat{\Delta}_a^{(P)}
	= \sum_{s \in C_a} w_{a,s}^{(P)}\, \hat{\Delta}_{a,s}.
\end{equation} 
The regression estimator can also be written in this form under additional assumptions on the outcome model, which are described in Section \ref{sec:reg}.

Equation \eqref{eq:generic} provides a useful interpretation of these estimators as plug-in estimators, in which  $\Delta_{a,s}$ is replaced by its respective period-specific estimator.  
The estimators below differ in (i) how $\hat{\Delta}_{a,s}$ is constructed (i.e., by using unadjusted, regression-adjusted, G-computation or augmented inverse probability weighting estimators -- we later mark these with a superscript to distinguish between them), (ii) which population is considered, which becomes evident in $w_{a,s}^{(P)}$, and (iii) which analysis dataset is used to fit any required model. To describe the role of the analysis dataset, let $\mathcal{I}_D$ denote the index set of participants used to fit the outcome model, with $D$ denoting the sample of the trial data we use. Let $\mathcal{I}_D^{(s)} = \{i \in \mathcal{I}_D : S_i = s\}$ be its period-$s$ subset.

\subsection{Unadjusted estimator}\label{sec:unadj} 

We can consider a covariate-unadjusted estimator that combines sample differences between treatment and control arm while weighting those differences according to the target population. For this, let
\begin{equation} \label{equnadj}
    \hat{\Delta}_{a,s}^{\,\mathrm{unadj}} = \bar{Y}_{a,s} - \bar{Y}_{0,s},
\end{equation}
where $\bar{Y}_{a,s}$ and $\bar{Y}_{0,s}$ denote the sample means in period $s$ for arm $a$ and control, respectively. Therefore,  $\hat{\Delta}_{a,s}^{\,\mathrm{unadj}}$ are the per-period differences in sample mean outcomes between arm $a$ and control. 

Thus the so-called unadjusted estimator is defined as:
\begin{equation}\label{eq:unadj}
	\hat{\Delta}_a^{(P)}
	= \sum_{s \in C_a} w_{a,s}^{(P)}\, \hat{\Delta}_{a,s}^{\,\mathrm{unadj}}.
\end{equation} 
This estimator does not account for covariates $X$, but by stratifying by period it avoids bias due to time trends.

\subsection{Regression estimator}\label{sec:reg} 

In the presence of time trends, a regression model can be used to obtain covariate-adjusted, period-specific contrasts. We consider two variants depending on whether non-concurrent control data are used.

\paragraph{Concurrent-controls (CC) model.}
Using data from $\mathcal{I}_{C_a} = \{i : S_i \in C_a\}$, that is, periods in which arm $a$ is active, the model is
\begin{equation}\label{eq-cc}
	g\left(E(Y_i \mid A_i, \, X_i,\, S_i)\right)
	= \alpha 
	+ \sum_{a'=1,2} \beta_{a'}\, I(A_i = a')
	+ \gamma^{\top} X_i
	+ \sum_{s \in C_a \setminus \{s_0\}} \tau_s\, I(S_i = s),
\end{equation}
where $s_0 = \min(C_a)$ is the reference period, $\beta_a$ is the conditional treatment effect of arm $a$ relative to control, and $\tau_s$ captures the period effect relative to $s_0$. The model is fitted by ordinary least squares for continuous outcomes with $g = \mathrm{id}$  or by maximum likelihood for binary outcomes with $g = \mathrm{logit}$.

\paragraph{Non-concurrent-controls (NCC) model.}
To borrow strength from non-concurrent control data, the model is fitted on $\mathcal{I}_{U_a} = \{i : S_i \in U_a\}$ (all periods up to and including $L_a$):
\begin{equation}\label{eq-ncc}
	g\left(E(Y_i \mid A_i,\, X_i,\, S_i)\right)
	= \alpha
	+ \sum_{a'=1,2} \beta_{a'}\, I(A_i = a')
	+ \gamma^{\top} X_i
	+ \sum_{s \in U_a \setminus \{s_0\}} \tau_s\, I(S_i = s),
\end{equation}
where $s_0 = \min(U_a)$. Here, $\tau_s$ quantifies the time drift over periods. The gain from including non-concurrent controls may arise because the additional data may improve estimation of the nuisance parameters in the model, particularly the period-effects $\tau_s$ and, when covariates are included, the prognostic effects $\gamma$. Improved estimation of the period effects yields a more precise estimate of the baseline control response, while improved estimation of $\gamma$ can further increase efficiency through more precise covariate adjustment.  

This requires an additional assumption: the period effects $\tau_s$ are common to all arms, i.e., there is no treatment-by-period interaction for any arm in the model, so that the control-arm time trend estimated from non-concurrent periods applies to the concurrent periods. The target estimand is unchanged, only the analysis set is enlarged. 
Other assumptions about the time trend can also be made (e.g., assuming linear time trends) and may be deemed more or less plausible depending on how well they capture the time trends in other treatment arms. Hence, model (\ref{eq-ncc}) mainly serves as a representative of time trend modelling used in Section \ref{sec:simcont} to quantify the efficiency gains due to stronger assumptions.

Let $\hat\Delta_a^{\mathrm{CC}}$ and $\hat\Delta_a^{\mathrm{NCC}}$ denote the estimators of $\beta_a$ obtained from the concurrent-controls model \eqref{eq-cc} and the non-concurrent controls model \eqref{eq-ncc}, respectively. 

Under correct model specification and the assumption of constant treatment effects across periods, both estimators target the conditional treatment effect $E(Y^{(a)} - Y^{(0)} \mid X, S)$.
Also note that in the linear case ($g = \mathrm{id}$), the conditional and marginal estimands coincide because $\beta_a = E(Y^{(a)}) - E(Y^{(0)})$.

In the linear case, the estimators of $\beta_a$ can be expressed as in equation \eqref{eq:generic}. When the model includes only treatment and period indicators (i.e., no additional covariates $X$), the adjusted within-period contrasts coincide with the unadjusted differences in means. In particular, for all
$s \in C_a$, $\hat{\Delta}_{a,s}^{\,\mathrm{CC}}=\hat{\Delta}_{a,s}^{\,\mathrm{unadj}}$. 
Thus, in this case, the concurrent-controls estimator can be written as:
\begin{align}\label{eq:reg-weighted}
	\hat\Delta_a^{\mathrm{CC}}
	&= \frac{\sum_{s \in C_a} \frac{n_{a,s}n_{0,s}}{m_{a,s}}\cdot \hat{\Delta}_{a,s}^{\mathrm{CC}}}
	{\sum_{s \in C_a} \frac{n_{a,s}n_{0,s}}{m_{a,s}}} 
\end{align} 

When $X$ is adjusted for, the OLS estimator remains a weighted average of within-period $X$-adjusted contrasts $\hat\Delta_{a,s}^{\,\mathrm{unadj}}$. On the other hand, the NCC estimator can in general not be written in the form of \eqref{eq:generic}. 
In the binary case ($g = \mathrm{logit}$), $\beta_a$ is a conditional log-odds ratio and $\hat\beta_a$ does not equal the marginal log-odds difference $\mathrm{logit}(E(Y^{(a)})) - \mathrm{logit}(E(Y^{(0)}))$ due to the non-linearity of the odds ratio \cite{daniel2021making}. This case is addressed separately in Section \ref{sec:binary}.

\subsection{Standardization (G-computation)}\label{sec:stand}

Regression standardization estimates the marginal treatment effect by averaging model-based predictions over a target population \citep{robins1986new, tsiatis2008covariate}. 
We fit an outcome model $\eta(A, S, X; \alpha, \beta, \gamma^{\top}, \tau^{\top})$ on the analysis set $\mathcal{I}_D$, with $g(\cdot)$ denoting its associated link function. This outcome model may be one of the models introduced in Section~\ref{sec:reg} or a different model chosen for this step. 

For each individual $i$, define the predicted potential outcome under treatment $a$ as:
\[
\hat{p}_i^a = g^{-1}\left(\eta(a, S_i, X_i;\hat\alpha, \hat\beta, \hat\gamma, \hat\tau)\right).
\]
The regression standardization estimator of the marginal effect of treatment $a$ relative to control $a=0$ is defined as
\begin{eqnarray}\label{eq:stand-pop}
	\hat{\Delta}_a^{(P),\mathrm{Stand}} &=& 
    \frac{1}{|\mathcal{I}_{(P)}|} \sum_{i\in \mathcal{I}_{(P)}}\left( \hat p_i^a - \hat p_i^0 \right)
\end{eqnarray}
where $\mathcal{I}_{(P)}$ is a sample from the target population, $|\mathcal{I}_{(P)}|$ denotes its size, and $\mathcal{I}_{(P)}$ may
differ from $\mathcal{I}_D$.

Two remarks are in order. First, the estimand targeted by \eqref{eq:stand-pop} depends on the covariate distribution of the target population represented by the sample $\mathcal{I}_{(P)}$ over which predictions are averaged:
changing this distribution shifts the estimator to a different causal parameter. Second, provided the outcome model is correctly specified, using a larger analysis set $\mathcal{I}_D$ that shares the same covariate distribution as the target population can improve precision in finite samples. Specifically, in finite samples, by leveraging additional data to better estimate the relationship between covariates and outcomes, this approach may reduce the variance of the predicted outcomes. For example, if the target population is the ACA population, the analysis set $\mathcal{I}_D$ used to fit the outcome model may be the \emph{concurrent assigned set} (ACA sample), the broader \emph{concurrent eligible set} (ECE sample), or the \emph{full trial data} (including non-concurrent patients); each choice involves a different bias-variance trade-off.

The regression standardization estimator can be written in the form \eqref{eq:generic} as follows. Define the period-specific standardized contrast
\begin{equation}\label{eq:stand-s}
	\hat{\Delta}_{a,s}^{\mathrm{Stand}}
	= \frac{1}{|\mathcal{I}_{(P)}^{(s)}|}
	\sum_{i \in \mathcal{I}_{(P)}^{(s)}}
	\bigl(\hat{p}_i^a - \hat{p}_i^0\bigr),
\end{equation}
where $\mathcal{I}_{(P)}^{(s)}$ is the index set of patients in population $P$ and period $s$.
Then the population-level standardization estimator for target population $(P)$ is
\begin{equation}\label{eq:stand}
	\hat{\Delta}_a^{(P),\mathrm{Stand}}
	= \sum_{s \in C_a} w_{a,s}^{(P)}\,
	\hat{\Delta}_{a,s}^{\mathrm{Stand}},
\end{equation}
where the weights $w_{a,s}^{(P)}$ are those defined in Section~\ref{sec:estimands}.  

\subsection{Augmented inverse probability weighting (AIPW)}
\label{sec:aipw}

The AIPW estimator augments standardization with an inverse probability weighting correction term, yielding a doubly robust estimator \cite{robins1994estimation}.

Let $\hat{p}_i^a$ be the predicted potential outcomes defined in Section~\ref{sec:stand}, obtained from the same outcome model $\eta(A, S, X;\hat\alpha,\hat\beta,\hat\gamma,\hat\tau)$, or another outcome model chosen for this step. 
Let $\pi_{a, (P)}(X_i, S_i) = P(A_i = a \mid X_i, S_i)$ be the probability of receiving treatment $a$ within the target population $P$, which is known by design in a randomized trial.
If randomization does not depend on $X$ and the allocation ratio is fixed within a period, then $\pi_{a, (P)}(X_i, S_i) = P(A_i = a \mid S_i)$ for all $i$. In this case, we denote it by $\pi_{a, (P)}(S_i)$, and we use this simpler notation from here on.


As the AIPW estimator relies on propensity scores, identification of $\Delta_a^{\mathrm{EP}}$ requires extrapolation of the treatment assignment mechanism to periods in which arm $a$ was not active \cite{santacatterina2025identification}. In these periods,
\[
P(A_i = a \mid X_i,S_i) = 0
\]
by design, since no participants could be allocated to arm $a$. 
Thus, estimation of $\Delta_a^{\mathrm{EP}}$ relies on extrapolation beyond the support of the observed treatment assignment mechanism. Since estimation of $\Delta_a^{\mathrm{EP}}$ would therefore require extrapolation beyond the support of the observed treatment assignment mechanism, we do not consider AIPW estimation for $\Delta_a^{\mathrm{EP}}$.

Define for each individual
\begin{equation}\label{eq:phi}
	\hat{\phi}_i^a
	= \frac{I(A_i = a)}{\pi_{a, (P)}(S_i)}
	\bigl(Y_i - \hat{p}_i^a\bigr)
	- \frac{I(A_i = 0)}{\pi_{0, (P)}(S_i)}
	\bigl(Y_i - \hat{p}_i^0\bigr)
	+ \hat{p}_i^a - \hat{p}_i^0.
\end{equation}
The period-specific AIPW contrast is
\begin{equation}\label{eq:aipw-s}
	\hat{\Delta}_{a,s}^{\mathrm{AIPW}}
	= \frac{1}{|\mathcal{I}_{(P)}^{(s)}|}
	\sum_{i \in \mathcal{I}_{(P)}^{(s)}} \hat{\phi}_i^a,
\end{equation}
and the population-level AIPW estimator is
\begin{equation}\label{eq:aipw}
	\hat{\Delta}_a^{(P),\mathrm{AIPW}}
	= \sum_{s \in C_a} w_{a,s}^{(P)}\,
	\hat{\Delta}_{a,s}^{\mathrm{AIPW}}.
\end{equation}
Note that the inverse-probability terms in \eqref{eq:phi} require $\pi^{(P)}_{a,s} > 0$ and $\pi^{(P)}_{0,s} > 0$ in every period contributing to the target population, which is the positivity assumption (iii) of Section \ref{sec:notation}. Although it holds by design for $s \in C_a$, AIPW can become unstable when one of these probabilities is close to zero.

The estimator \eqref{eq:aipw} is doubly robust: it is a consistent estimator of $\Delta_a^{(P)}$ if either the outcome model $\eta(a,s,x;\hat\beta,\hat \gamma, \hat \tau)$ or the propensity score model $\pi_a(S_i)$ is correctly specified. Because $\pi_a(S_i)$ is known by design in a randomized trial, double robustness implies that  $\hat{\Delta}_a^{(P),\mathrm{AIPW}}$ is always consistent regardless of whether the outcome model is correctly specified. 

In platform trials, periods are defined as intervals during which the set of available arms remains fixed (i.e., no arm enters or leaves within a period) and during which the randomization ratio remains unchanged. Provided randomization is not covariate-adaptive in this work, this implies that the randomization probability $\pi_a(X_i, S_i)$ is constant within each period. Under this simplification, the mean of \eqref{eq:phi} simplifies, and the AIPW estimator reduces to the standardization estimator when canonical generalized linear models are used for the outcome models and include treatment-by-period interactions. 

Importantly, the propensity score $\pi_a(S_i)$ depends not only on the period but also on the target population $(P)$, since the relevant risk set changes across populations. Consider the comparison of treatment $a=2$ versus control in the trial of Figure~\ref{fig:example}. For the ACA population, only patients assigned to arm $2$ or control are included, so the propensity score is
$$\pi_a^{\mathrm{ACA}}(S_i = s) = \frac{n_{a,s}}{n_{0,s} + n_{2,s}} 
= \frac{n_{a,s}}{m_{2,s}}, \qquad s \in C_2,$$ 
which is constant across periods. For the ECE population, patients randomised to other active experimental arms are also included, so the denominator expands to cover all arms available in each period:
$$
    \pi_a^{\mathrm{ECE}}(S_i = 2) = \frac{n_{a,2}}{n_{0,2} + n_{1,2} + n_{2,2}}
    = \frac{n_{a,2}}{N_2},
    \qquad
    \pi_a^{\mathrm{ECE}}(S_i = 3) = \frac{n_{a,3}}{n_{0,3} + n_{2,3}}
    = \frac{n_{a,3}}{N_2}.
$$
In period 2, where arm $1$ is still active, the ECE and ACA propensity scores differ. But in period 3 they coincide because arm $1$ has been discontinued. More generally, $\pi_a^{\mathrm{ECE}}(s) = 
n_{a,s}/N_s$ and $\pi_a^{\mathrm{ACA}}(s) = n_{a,s}/m_{a,s}$, 
with the two agreeing whenever no other experimental arm is active in period $s$.

\section{Analytical results of consistency and variance for continuous endpoints}\label{sec:rescont}

To study differences between estimands and between estimators, we consider the following data-generating model: 
\begin{equation}\label{datagen-model}
	Y_i(a) = \alpha_{S_i} + \theta_a + \phi_{a,S_i}
	+ \bigl(\beta + \kappa_{S_i} + \psi_a\bigr) X_i + \varepsilon_i(a),
\end{equation}
where $X_i$ is continuous and with
\[
X_i \mid S_i = s \sim F_s,
\qquad
\varepsilon_i(a) \mid S_i = s \sim N(0,\sigma_{a,s}^2).
\]
The terms can be interpreted as follows: $\alpha_s$ is the period effect,
$\theta_a$ the effect of treatment,
$\phi_{a,s}$ the treatment-period interaction,
$\beta$ the effect of covariate, 
$\kappa_s$ the covariate-period interaction,
$\psi_a$ is treatment-covariate interaction, and
$F_s$ is the period-specific covariate distribution.

The analytical results are organised as follows. In Section \ref{sec:estimands_coincide}, we establish under which conditions concurrent estimands coincide, determining then when the choice of target population is crucial. 
In Section \ref{sec:consistency_estimators}, we examine the consistency of the estimators. In Section \ref{sec:superpopulation}, we decompose the superpopulation variance  into three sources of variability: the variability from conditioning on $X$ and $S$, the variability driven by the covariates $X$ and the variability arising from the periods $S$.  
In Section \ref{sec:condvar}, we compare the variance conditioning on $X$ and $S$ according to the estimand. In Section \ref{sec:condvarunadj}, we compare the variance according to the estimand when using unadjusted estimators, and, finally, in Section \ref{sec:exact-aipw}, we examine the variance for the AIPW estimator.   

\subsection{Situations in which the concurrent estimands coincide}\label{sec:estimands_coincide}

Under the model in \eqref{datagen-model}, the period-specific treatment effect is
\[
\Delta_{a,s} = E\{Y_i(a) - Y_i(0) \mid S_i=s\}
= \theta_a + \phi_{a,s} + \psi_a \cdot E(X \mid S=s).
\]
Hence, the concurrent estimands ECE, ACA, and LACA coincide for every trial design, that is, whatever the period sizes, allocation ratios and entry and exit times of competing arms, if and only if the period-specific effects are constant over $s \in C_a$. 
Within model \eqref{datagen-model}, $\Delta_{a,s}$ is constant over $s \in C_a$ if
\[\phi_{a,s} \text{ is constant in } s, \text{ i.e., }
\phi_{a,s} = \phi_{a}  \text{ for all } s \in C_a,
\]
and either
$$
\psi_a = 0, \ \ \text{or} \ \ 
E(X \mid S=s) \text{ is constant in } s.
$$ 
If the $\Delta_{a,s}$ vary, the estimands generally differ and can agree only through particular features of the design. For example, ECE and ACA coincide whenever $N_s/m_{a,s}$ is constant across $s \in C_a$, because their weights are then identical. 

\subsection{Consistency of the estimators} \label{sec:consistency_estimators}

For estimators that can be written as a weighted average of period-specific estimators, as in \eqref{eq:generic}, the consistency of $\hat{\Delta}_a^{(P)}$ follows from the consistency of the period-specific estimators $\hat{\Delta}_{a,s}$ for $\Delta_{a,s}$.  Therefore, $\hat{\Delta}_a^{(P)}$ is asymptotically a fixed weighted average of the consistent estimators $\hat{\Delta}_{a,s}$, implying consistency of the overall estimator.

The period-specific AIPW estimators $\hat\Delta_{a,s}^{\mathrm{AIPW}}$ are consistent for $\Delta_{a,s}$ whenever the randomization probabilities are correctly specified -- which holds by design in a randomized trial -- regardless of whether the outcome model is correctly specified. Consequently, $\hat\Delta_a^{(P),\mathrm{AIPW}}$ is consistent for every target population $(P)$.

For standardization (G-computation) and for linear regression, consistency of $\hat\Delta_{a,s}$ depends on correct specification of the outcome model $E(Y \mid A, S, X)$.  

If the outcome model is correctly specified, $\hat\Delta_{a,s}^{\mathrm{Stand}}$ is consistent in every period, and hence $\hat\Delta_a^{(P),\mathrm{Stand}}$ is consistent for every estimand $(P)$. 

A model that includes treatment-by-period interactions is saturated in $(A,S)$ and so imposes no restriction on how the treatment effect varies across periods, although its covariate part may still be misspecified.
If it is a canonical generalized linear model (e.g.\ identity or logit link) fitted by maximum likelihood on a dataset containing all arm-$a$ and control participants of the target sample, its residuals sum to zero within each treatment-period cell. Because the randomization probabilities are constant within periods, the augmentation term of AIPW then vanishes, so the standardization estimator is numerically equivalent to AIPW and takes its consistency even when the covariate part of the model is misspecified. 
If no baseline covariates are included, this estimator reduces to the unadjusted per-period difference in means.

When treatment-by-period interactions are omitted and the model is otherwise misspecified, $\hat\Delta_{a,s}^{\mathrm{Stand}}$ may be biased for $\Delta_{a,s}$ in every individual period. Even so, $\hat\Delta_a^{(P)}$ can remain consistent, because the period-specific biases can cancel in the weighted sum whenever the implicit weighting of the (misspecified) model matches the target weights $w_{a,s}^{(P)}$. For a model fitted on concurrent data, this gives the following sufficient conditions:
\begin{itemize} 
\item For ACA, consistency holds if the ratio $n_{a,s}/m_{a,s}$ is constant across periods in $C_a$, a condition often satisfied in practice.  
\item For ECE, consistency requires that the ratio $n_{a,s}/N_s$ be constant across periods in $C_a$, which is typically violated unless the number of treatment arms does not change over time. 
\end{itemize} 

Notably, in these special cases the overall estimator may remain consistent even if the individual period-specific estimators $\hat{\Delta}_{a,s}$ are not.

\subsection{Conditional-on-$X,S$ variance and superpopulation variance} \label{sec:superpopulation}

To compare the precision for different estimands, we examine the variance of estimators for population-level estimands $\Delta_a^{(P)}$. 

The variance of $\hat\Delta_a^{(P)}$ can be decomposed into a design-based component, obtained by conditioning on the realized covariates $X$ and period composition $S$, and two further components that arise only when $X$ and $S$ are considered as random draws from a superpopulation.  
\begin{equation}\label{eq:var-decomp}
	\mathrm{Var}\!\left(\hat\Delta_a^{(P)}\right)
	= \underbrace{E_S\!\left[E_X\!\left[\mathrm{Var}\!\left(\hat\Delta_a^{(P)}\mid X,S\right)\right]\right]}_{\text{(i) design-based variance}}
	+\underbrace{E_S\!\left[\mathrm{Var}_X\!\left(E\!\left[\hat\Delta_a^{(P)}\mid X,S\right]\right)\right]}_{\text{(ii) random-}X\text{ component}}
	+\underbrace{\mathrm{Var}_S\!\left(E_{X}\!\left[\hat\Delta_a^{(P)}\mid S\right]\right)}_{\text{(iii) random-}S\text{ component}}.
\end{equation}

To characterise each term, note that under the data-generating model \eqref{datagen-model} the conditional expectation given both $X$ and $S$ is
\begin{equation}\label{eq:cond-exp-XS}
	E\!\left[\hat\Delta_a^{(P)} \mid X, S\right]
	= \sum_{s \in C_a} w_{a,s}^{(P)}\!\left(\theta_a + \phi_{a,s} + \psi_a \bar{X}_s\right),
\end{equation}
where $\bar{X}_s$ is the realized sample mean of $X$ in period $s$. The conditional expectation given $S$ alone is
\begin{equation}\label{eq:cond-exp-S}
	E_X\!\left[\hat\Delta_a^{(P)} \mid S\right]
	= \sum_{s \in C_a} w_{a,s}^{(P)}\!\left(\theta_a + \phi_{a,s} + \psi_a\, E(X \mid S=s)\right).
\end{equation}

The three terms in \eqref{eq:var-decomp} can now be interpreted as follows.

\begin{itemize}
	
	\item \textbf{Term (i)} refers to the design-based variance. This is the quantity computed in Section~\ref{sec:condvar}, $\sum_{s \in C_a} (w_{a,s}^{(P)})^2\, v_{a,s}$, now understood as averaged over all possible realisations of $X$ and $S$. It captures the variance due to random treatment assignment, holding the covariates and period fixed.
	
	\item \textbf{Term (ii)} refers to the random-$X$ component. This term is the variance of expression \eqref{eq:cond-exp-XS} over the random draw of $X$ within fixed $S$, averaged over $S$. From \eqref{eq:cond-exp-XS}, this expression depends on $X$ only through the within-period covariate means $\bar{X}_s$. Hence, Term (ii) is non-zero if and only if $\psi_a \neq 0$, i.e., there is a treatment--covariate interaction. When $\psi_a = 0$, expression \eqref{eq:cond-exp-XS} does not depend on $X$ and this term vanishes.
	
	\item \textbf{Term (iii)} refers to the random-$S$ component. This term is the variance of expression \eqref{eq:cond-exp-S} over the random realisation of $S$. From \eqref{eq:cond-exp-S}, this expression varies with $S$ through: (a) if $\phi_{a,s} \neq 0$ (treatment--period interaction), different period realisations produce different period-specific treatment effects, even if $X$ plays no role; (b) if $\psi_a \neq 0$ and the covariate distribution $F_s$ varies across periods, the period-specific mean $E(X \mid S=s)$ changes with $s$, adding a further source of variability through the interaction between the treatment--covariate effect and the shifting covariate mix. Additionally, if the weights $w_{a,s}^{(P)}$ are themselves random because they depend on realised per-period sample sizes, this introduces yet another source of variation.
	
\end{itemize}

The superpopulation variance captures the variability induced by redrawing the covariates (term (ii)) and the period composition (term (iii)). Importantly, note that terms (ii) and (iii) vanish simultaneously under the same conditions that make the concurrent estimands coincide, that is to say: when $\psi_a = 0$ and the period-specific effects $\Delta_{a,s}$ are constant across $s\in C_a$, superpopulation variance and design-based variance are equal.

For a given trial design, terms (ii) and (iii) represent additional variance components that are typically ignored in practice and which can matter when effects are heterogeneous across periods.

\subsection{Variance formula for treatment effect estimators} \label{sec:condvar}

Here, we consider the variance conditioning on $X$ and $S$, that is, the design-based variance (term (i) in \eqref{eq:var-decomp}). To do so, let
\[
\hat\Delta_a^{(P)} = \sum_{s \in C_a} w_{a,s}^{(P)} \hat\Delta_{a,s},
\]
where $\hat\Delta_{a,s}$ is an unbiased estimator of $\Delta_{a,s}$ with $\mathrm{Var}(\hat\Delta_{a,s}) = v_{a,s}$, 
and assume independence across periods. Then
\[
E\bigl(\hat\Delta_a^{(P)}\bigr) = \Delta_a^{(P)} = \sum_{s \in C_a} w_{a,s}^{(P)} \Delta_{a,s},
\]
and
\begin{equation}\label{vareq}
	\mathrm{Var}\bigl(\hat\Delta_a^{(P)}\bigr) = \sum_{s \in C_a} \bigl(w_{a,s}^{(P)}\bigr)^2 v_{a,s}.
\end{equation}

If the period-specific effects are constant across $s$, estimands coincide as seen before but the estimands coincide and the estimators differ only in variance. Because the weights $w_{a,s}^{(P)}$ differ from populations, the estimators combine period-specific information differently and therefore have different sampling variances. 

Note that the variance formula does not require the $\Delta_{a,s}$ to be constant across $s$. Hence, the comparisons in variance continue to hold when the estimands differ across populations. 
Therefore, we can use formula \eqref{vareq} to compare the precision for different estimands, even when estimands do not have a common target.  

\subsection{Variance comparison using unadjusted estimators under some design conditions} \label{sec:condvarunadj}  

To derive explicit variance ratios between estimands, we introduce two conditions:
\begin{itemize}
	\item Condition A: $\sigma_{a,s}=\sigma$ for all $a,s$  (constant residual variance across arms and periods).
    \item Condition B: $\pi_{a,s}=\pi_a$ for all $s\in C_a$, where $\pi_{a,s}=n_{a,s}/m_{a,s}$ (constant allocation ratio across periods).
\end{itemize} 

Consider the per-period unadjusted estimator, $\hat\Delta_{a,s}^{\,\mathrm{unadj}}$, defined in \eqref{equnadj}. With Condition A,  the variance of the unadjusted difference in means within period $s$ simplifies to
\begin{equation}\label{eq:vas-condA}
v_{a,s} = \sigma^2 \left(\frac{1}{n_{a,s}}+\frac{1}{n_{0,s}}\right) =\frac{\sigma^2}{m_{a,s} \pi_{a,s} (1-\pi_{a,s})} 
\end{equation}

Under Conditions A and B, the period-specific variance simplifies to
\begin{equation}\label{eq:vas-simplified}
    v_{a,s} = \frac{\sigma^2}{\pi_a(1-\pi_a)\,m_{a,s}} 
    = \frac{c_a}{m_{a,s}}, 
    \qquad c_a = \frac{\sigma^2}{\pi_a(1-\pi_a)}.
\end{equation}

\paragraph{Regression estimator's weights.}
With \eqref{eq:vas-condA}, we note that the CC regression estimator in \eqref{eq:reg-weighted} targets the inverse-variance-weighted estimand, which equals ACA only under Condition B and equals neither ECE nor ACA otherwise.

\paragraph{ACA achieves minimum variance.}
The inverse-variance optimal weights are $w_{a,s}^\star \propto 1/v_{a,s} = m_{a,s}/c_a \propto m_{a,s}$, which coincide exactly with $w_{a,s}^{\mathrm{ACA}}$. Hence, under Conditions A and B, the ACA estimator achieves minimum variance among all estimators of the form $\sum_{s \in C_a} w_{a,s} \hat\Delta_{a,s}^{\mathrm{unadj}}$ with $\sum_s w_{a,s} = 1$. Both the ECE and LACA estimators therefore have variance at least as large as ACA.

If Condition A holds but Condition B fails, the optimal weights are $w_{a,s}^\star \propto \pi_{a,s}(1-\pi_{a,s})m_{a,s}$, which up-weight periods with more balanced allocation. The ACA weights $w_{a,s}^{\mathrm{ACA}} \propto m_{a,s}$ no longer coincide with the optimal weights, as they ignore the true allocation ratios $\pi_{a,s}$. The mismatch arises because periods with a more balanced allocation contribute more information per patient than periods with unequal splits, but ACA ignores this.

\paragraph{LACA versus ACA.}
Since $\mathrm{Var}(\hat\Delta_a^{\mathrm{LACA}}) = v_{a,L_a} = c_a/m_{a,L_a}$ and $\mathrm{Var}(\hat\Delta_a^{\mathrm{ACA}}) = c_a/\sum_{s \in C_a} m_{a,s}$,
\begin{equation}\label{eq:laca-aca-ratio}
    \frac{\mathrm{Var}(\hat\Delta_a^{\mathrm{LACA}})}
    {\mathrm{Var}(\hat\Delta_a^{\mathrm{ACA}})} 
    = \frac{\sum_{s \in C_a} m_{a,s}}{m_{a,L_a}} \geq 1,
\end{equation}
with equality only if $|C_a|=1$.  The ratio grows with the size of earlier concurrent periods relative to the last one. 
The efficiency loss of LACA over ACA is largest when substantial information accumulated in earlier periods is discarded.

\paragraph{ECE versus ACA.}
Substituting \eqref{eq:vas-simplified} into the general variance formula \eqref{vareq} gives
\begin{align}
    \mathrm{Var}(\hat\Delta_a^{\mathrm{ACA}}) 
    &= \frac{c_a}{\displaystyle\sum_{s \in C_a} m_{a,s}}, \label{eq:var-aca} \\[6pt]
    \mathrm{Var}(\hat\Delta_a^{\mathrm{ECE}}) 
    &= c_a\,
    \frac{\displaystyle\sum_{s \in C_a} N_{s}^2/m_{a,s}}
    {\displaystyle\Bigl(\sum_{s \in C_a} N_{s}\Bigr)^2}. \label{eq:var-ece}
\end{align}
It follows that $\mathrm{Var}(\hat\Delta_a^{\mathrm{ECE}}) \geq \mathrm{Var}(\hat\Delta_a^{\mathrm{ACA}})$, with equality if and only if $N_{s}/m_{a,s}$ is constant across $s \in C_a$, which holds only when no competing arm is active or when entries and exits of competing arms happen to preserve the proportion $m_{a,s}/N_{s}$ in every period.

For the case $|C_a|=2$, defining $R_N = N_{1}/N_{2}$ and $R_m = m_{a,1}/m_{a,2}$, the ratio simplifies to
\begin{equation}\label{eq:var-ratio-ece-aca}
    \frac{\mathrm{Var}(\hat\Delta_a^{\mathrm{ECE}})}{\mathrm{Var}(\hat\Delta_a^{\mathrm{ACA}})} 
    = 1 + \frac{(R_N - R_m)^2}{R_m(1+R_N)^2}.
\end{equation}
The ratio exceeds 1 whenever $R_N \neq R_m$ and grows with the discrepancy between the two ratios. 

\subsection{Exact distribution of the adjusted estimators in the Gaussian linear model} \label{sec:exact-aipw}

The closed-form variances above concern the unadjusted estimator. For the covariate-adjusted estimators, the per-period contrasts share a single fitted model and are no longer independent across periods, so the simple form \eqref{vareq} does not apply. In the Gaussian linear model, however, an exact finite-sample distribution is available, which we now derive. 

For this, we assume a continuous outcome, a linear working outcome model fitted by generalized least squares (GLS), and Gaussian errors $\varepsilon\mid\mathcal X\sim N(0,\bm\Sigma)$ with $\bm\Sigma$ block-diagonal ($\sigma^2_{a,s}I_{n_{a,s}}$ blocks). We assume  that randomization probabilities are known by design.  

Let \[
\mathbf Y
=
\begin{pmatrix}
\mathbf Y_{0,1}^{\top}&
\mathbf Y_{1,1}^{\top}&
\mathbf Y_{0,2}^{\top}&
\mathbf Y_{1,2}^{\top}&
\mathbf Y_{2,2}^{\top}&
\mathbf Y_{0,3}^{\top}&
\mathbf Y_{2,3}^{\top}
\end{pmatrix}^{\top},
\]
be the outcomes by treatment-period cell where \(\mathbf Y_{a,s}\) denotes the vector of observed outcomes in treatment-period cell \((a,s)\). The  model can be written as $\mathbf Y=\mathcal X\vartheta+\varepsilon$,
where $\mathcal X$ is the design matrix induced by the realized cells and covariate values and $\vartheta$ collects the period, treatment, treatment-period, covariate, covariate-period and treatment-covariate parameters of \eqref{datagen-model}. The GLS fit $\hat\vartheta=G\mathbf Y$, with $G=V_\vartheta\mathcal X^\top\bm\Sigma^{-1}$ and $V_\vartheta=(\mathcal X^\top\bm\Sigma^{-1}\mathcal X)^{-1}$, is linear in $\mathbf Y$; hence so
are the fitted potential-outcome predictions and the AIPW pseudo-outcomes. 

Collecting terms, the population-level AIPW estimator is a single linear contrast
\begin{equation}\label{eq:aipw-linear}
  \hat\Delta_a^{(P),\mathrm{AIPW}} = \ell_a^{(P)\top}\mathbf Y,
\end{equation}
and, conditional on the design,
\begin{equation}\label{eq:aipw-exact}
  \hat\Delta_a^{(P),\mathrm{AIPW}}\mid\mathcal X
  \sim N\!\left(\ell_a^{(P)\top}\mathcal X\vartheta,\;
  \ell_a^{(P)\top}\bm\Sigma\,\ell_a^{(P)}\right).
\end{equation}
The explicit form of $\ell_a^{(P)}$ is given in Appendix \ref{app:der}. 
Equation \eqref{eq:aipw-exact} is the design-based (conditional on $X,S$) counterpart of \eqref{vareq} for the adjusted estimator: the exact conditional variance is $\ell_a^{(P)\top}\bm\Sigma\,\ell_a^{(P)}$, with no independence-across-periods assumption.

\paragraph{Standardization is the BLUE; AIPW is weakly less efficient under correct specification.} 

Let $H_a^{(P)}$ be the
design matrix for participants in the target-population sample
$\mathcal I_{(P)}$, constructed after setting treatment to $a$ for
every participant while retaining their observed periods and
covariates. Define $H_0^{(P)}$ analogously by setting treatment to
control. For participant $i$ in period $s$, let
\[
q_{a,i}^{(P)}
=
\frac{w_{a,s}^{(P)}}{|\mathcal I_{(P)}^{(s)}|}.
\]
The corresponding model-based contrast is
\[
\Delta_a^{(P)}(\mathcal X)
=
q_a^{(P)\top}
\left(H_a^{(P)}-H_0^{(P)}\right)\vartheta.
\]
The standardization estimator replaces $\vartheta$ by its GLS
estimator:
\[
\hat\Delta_a^{(P),\mathrm{Stand}}
=
q_a^{(P)\top}
\left(H_a^{(P)}-H_0^{(P)}\right)\hat\vartheta
=
c_a^{(P)\top}\hat\vartheta,
\qquad
c_a^{(P)}
=
\left(H_a^{(P)}-H_0^{(P)}\right)^\top q_a^{(P)}.
\]
Under the correctly specified linear model with known covariance
matrix, this is the GLS BLUE of the conditional model-based contrast,
with variance
\[
c_a^{(P)\top}V_\vartheta c_a^{(P)}.
\]

The AIPW estimator \eqref{eq:aipw-linear} is also a linear unbiased estimator of the same contrast when the outcome model is correctly specified, but it is generally not the GLS solution. By the Gauss--Markov theorem, 
\begin{equation}\label{eq:gm-ordering}
  \ell_a^{(P)\top}\bm\Sigma\,\ell_a^{(P)} \;\ge\; c_a^{(P)\top}V_\vartheta\,c_a^{(P)},
\end{equation}
with equality in special cases. In particular, when the outcome model includes treatment-by-period interactions and the randomization probabilities are constant within periods, AIPW coincides with standardization. 

Thus, when the outcome model is correctly specified, AIPW cannot be more efficient than GLS standardization within the class of linear unbiased estimators. The efficiency loss may be negligible in some settings but can be substantial under highly unequal randomization probabilities.

\paragraph{Roles of the target population and the analysis dataset.}
Under the additive linear working model, which contains neither
treatment-by-period nor treatment-by-covariate interactions, the
predicted treatment contrast is
\[
\hat p_i^a-\hat p_i^0=\hat\beta_{a,D}
\]
for every participant, where $D$ denotes the analysis dataset used to
fit the model. Because the target-population weights sum to one,
\[
\hat\Delta_{a,D}^{(P),\mathrm{Stand}}
=
\hat\beta_{a,D}.
\]
Consequently, for a fixed analysis dataset, the regression and
standardization estimators and their variances do not depend on the
target population $(P)$.

The analysis dataset does matter. Fitting the correctly specified GLS
model to a broader dataset can improve estimation of the shared period
and covariate effects and thereby reduce the variance of
$\hat\beta_{a,D}$. Thus, fitting on the ECE dataset may be more precise
than fitting on ACA, while using non-concurrent data may provide
additional precision under the required extrapolation assumptions.
This target-population invariance does not generally extend to AIPW,
because its residual-correction term can depend on the target sample
and its weights.

\section{Simulations with a continuous endpoint}\label{sec:simcont}

To evaluate bias and variance between estimators for each estimand, we conducted a simulation. 

We simulated three-period trials with two experimental arms and a shared control, as illustrated in Figure \ref{fig:example}.
To simulate data, we considered the data-generating model in \eqref{datagen-model}.  For each scenario, we simulated 10000 trials and estimated the estimands for the populations ACA, ECE, and LACA, using the unadjusted estimator, linear model (LM), G-computation, and AIPW (as explained in Section \ref{sec:estimators}). For the latter, we considered estimation using different analysis datasets.    

The considered scenarios are described below in terms of the parameters of the data-generating model in \eqref{datagen-model}. Scenarios 1-6 mimic, for the adjusted estimators, the settings analysed in Section \ref{sec:condvarunadj} for the unadjusted estimator; Scenarios 7-9 introduce treatment-period and treatment-covariate interactions and two extreme-allocation designs. 
We considered scenarios under the global null hypothesis ($\theta_a = 0$ for all $a$) and scenarios under the alternative hypothesis with $\theta_1 = 1$, $\theta_2 = 2$. 
In all scenarios, the residual variance is constant, $\sigma_{a,s}=\sigma=1$, satisfying Condition A of Section \ref{sec:condvarunadj}. The covariate distribution is the same across periods, $F_s = F$ for all $s$, unless stated otherwise. 
The sample sizes per period are equal, $N_s = 100$ for all $s$, unless stated otherwise. 
In scenarios with time trends, we simulated stepwise time trends such that the period effects increase by 1 from period to period, that is, $\alpha_s=s-1$ for $s=1,2,3$.

Scenarios are described in what follows and summarised in Table \ref{tab:scenarios}:
\begin{enumerate}

\item \textbf{Scenario 1: Equal allocation, no trend, no covariate effects.} 
Global null hypothesis, there are no treatment-period interactions ($\phi_{a,s} = 0$), no covariate effects ($\beta = \kappa_s = \psi_a = 0$), and no time trend ($\alpha_s = 0$).  Allocation is equal in all periods ($\pi_{a,s} = \pi_a$ for all $a$ and $s$, satisfying Condition B), and period sample sizes are equal.

\item \textbf{Scenario 2: Unequal allocation, no trend.}
Alternative hypothesis, but there are no treatment-period interactions, no covariate effects, and no time trends ($\phi_{a,s} = 0$, $\beta = \kappa_s = \psi_a = 0$, $\alpha_s = 0$).  Period sizes are equal.
The allocation ratio varies across periods ($2:1$; $2:1:1$; $2:1$) but Condition B still holds. 

\item \textbf{Scenario 3: Equal allocation, time trend, unequal period sizes.}
Alternative hypothesis; also, there is a stepwise time trend ($\alpha_s=s-1$ for $s=1,2,3$), and no covariate effects ($\beta = \kappa_s = \psi_a = 0$, $\phi_{a,s} = 0$).  Allocation is equal within each period, but period sizes differ ($N_1 = 50$, $N_2 = 150$, $N_3 = 100$).  

\item \textbf{Scenario 4: Equal allocation, no trend, strong covariate effect.}
As in Scenario 1, except $\beta = 2$, so the covariate $X_i$ is a strong predictor of $Y_i(a)$.  All other interaction terms remain zero ($\kappa_s = \psi_a = \phi_{a,s} = 0$) and $\alpha_s = 0$. Conditions~A and~B are both satisfied.  

\item \textbf{Scenario 5: Equal allocation, time trend, unequal period sizes, strong covariate effect.}
Combines the design of Scenario~3 (alternative hypothesis,  stepwise time trend, unequal period sample sizes and equal allocation) with the covariate effect of Scenario 4 ($\beta = 2$).  All interaction terms remain zero ($\kappa_s = \psi_a = \phi_{a,s} = 0$). Conditions A and B are satisfied.   

\item \textbf{Scenario 6: Equal allocation, time trend, treatment--period interaction.}
Alternative hypothesis; also, there is a stepwise time trend and a treatment-period interaction for arm 1 ($\phi_{1,s} = 3$ for the period in which arm 1 competes with arm 2, while $\phi_{2,s} = 0$).  No covariate effects are present ($\beta = \kappa_s = \psi_a = 0$).  Allocation is equal and period sizes are equal. 

\item \textbf{Scenario 7: Treatment--covariate interaction with covariate shift across periods.} 
Alternative hypothesis with a covariate is a strong predictor ($\beta = 2$). There is treatment-covariate interaction  for arm 1 ($\psi_1 = 2$) but not for arm 2 ($\psi_2 = 0$), 
and the covariate distribution shifts across periods: $X$ is drawn from a two-component mixture of $N(0,0.5^2)$ and $N(3,0.5^2)$, with the weight on the high-mean component increasing over periods (80\%/20\%, 40\%/60\%, and 20\%/80\% low/high in periods 1, 2, and 3, respectively). This yields period-specific means $E(X\mid S=1)=0.6$, $E(X\mid S=2)=1.8$, $E(X\mid S=3)=2.4$, i.e.
the covariate distribution differs across periods ($F_s \neq F_{s'}$), but the covariate-outcome slope itself does not ($\psi_a = 0$ and $\kappa_s = 0$). There is no treatment--period interaction ($\phi_{a,s}=0$) and no time trend ($\alpha_s = 0$); period sizes are equal.

\item \textbf{Scenario 8: Dominant competing arm.}
Alternative hypothesis with no	time trend, no covariate effect and no interaction ($\alpha_s = \beta = \kappa_s = \psi_a = \phi_{a,s} = 0$). Period sizes are highly unequal ($N_1 = 200$, $N_2 = 1800$, $N_3 = 200$), and in period 2 arm 2 absorbs almost all patients (allocation $0.98$ to arm 2, $0.01$ each to control and arm 1). 

\item \textbf{Scenario 9: Highly variable allocation ratio.}
Alternative hypothesis with no time trend, no covariate effect, and no interaction
($\alpha_s = \beta = \kappa_s = \psi_a = \phi_{a,s} = 0$). Period sample sizes are moderately unequal ($N_1 = 400$, $N_2 = 200$, $N_3 = 200$). The allocation ratio varies strongly across periods for the comparison arm 1 versus control:  period 1 allocates $0.95$ to control and only $0.05$ to arm 1, while periods 2 and 3 use balanced allocation ($1:1:1$ and $1:1$, respectively). Note that by contrast, the control vs arm 2 allocation remains balanced ($0.5$) in both periods in which arm 2 is active. In this scenario then Condition~B is strongly violated.

\end{enumerate}

For each scenario, we evaluated the comparisons arm 2 versus control and arm 1 versus control. For each, we computed the mean, standard deviation and median of the estimator over the simulated trials, and also the mean, standard deviation, and median of the variance and the standard error of the model-based estimator. To compute the model-based variance estimator, we used classical homoscedastic OLS variance for the LM,  also classical OLS via delta method for G-computation and empirical influence-function variance for the AIPW. We also reported the empirical variance of the estimator by computing the variance of the estimates over the simulated trials per scenario.    

Throughout, the reported Monte Carlo standard deviations and the estimated standard errors target the superpopulation variance \eqref{eq:var-decomp}, since each replicate redraws the covariates, the treatment assignment and the errors (period sizes are held fixed by design). This coincides with the design-based variance of Section \ref{sec:condvar} only when the treatment effect is homogeneous (meaning no treatment-period or treatment-covariate interaction), where terms (ii) and (iii) vanish. In Scenarios with treatment effect heterogeneity (Scenarios 6-7), term (iii) is zero as period sample sizes are considered fixed, and term (ii) only remains in Scenario 7 (where $\psi\neq0$); thus, the reported variances additionally reflect term (ii) in Scenario 7.  

\begin{table}[h!]
\centering
\caption{Summary of data-generating and design parameters used in each simulation scenario. $\theta_1,\theta_2$ are treatment effects; $\alpha_s=(\alpha_1,\alpha_2,\alpha_3)$ the period effects; $\beta$ the baseline covariate effect; $\psi_1,\psi_2$ the treatment-covariate interactions; $\phi_{a,s}$ the treatment-period interactions. Allocation ratios are given as control:arm~1:arm~2 for periods 1; 2; 3 (inactive arms omitted). ``Cond.\ B'' reports whether $n_{a,s}/m_{a,s}$ is constant across $s\in C_a$, separately for the arm-1-vs-control and arm-2-vs-control contrasts. Condition~A ($\sigma_{a,s}=\sigma=1$) and $\kappa_s=0$ hold throughout.}
\label{tab:scenarios}
\resizebox{\textwidth}{!}{%
\begin{tabular}{clccccccccc}
\toprule
Scenario & Hyp. & $\theta_1,\theta_2$ & $\alpha_s$ & $\beta$ & $\psi_1,\psi_2$ & $\phi_{a,s}$ & $N_1,N_2,N_3$ & Allocation (period 1; 2; 3) & Cond.\ B (Arm1/Arm2) & $F_s$ \\
\midrule
1 & $H_0$ & $0,0$       & $0,0,0$ & $0$ & $0,0$ & $0$               & $100,100,100$ & $1{:}1$; $1{:}1{:}1$; $1{:}1$       & Y / Y          & $N(0,1)$ \\
2 & $H_1$ & $1,2$       & $0,0,0$ & $0$ & $0,0$ & $0$               & $100,100,100$ & $2{:}1$; $2{:}1{:}1$; $2{:}1$       & Y$^{b}$ / Y$^{b}$ & $N(0,1)$ \\
3 & $H_1$ & $1,2$       & $0,1,2$ & $0$ & $0,0$ & $0$               & $50,150,100$  & $1{:}1$; $1{:}1{:}1$; $1{:}1$       & Y / Y          & $N(0,1)$ \\
4 & $H_0$ & $0,0$       & $0,0,0$ & $2$ & $0,0$ & $0$               & $100,100,100$ & $1{:}1$; $1{:}1{:}1$; $1{:}1$       & Y / Y          & $N(0,1)$ \\
5 & $H_1$ & $1,2$       & $0,1,2$ & $2$ & $0,0$ & $0$               & $50,150,100$  & $1{:}1$; $1{:}1{:}1$; $1{:}1$       & Y / Y          & $N(0,1)$ \\
6 & $H_1$ & $1,2$       & $0,1,2$ & $0$ & $0,0$ & $\phi_{1,2}{=}3$  & $100,100,100$ & $1{:}1$; $1{:}1{:}1$; $1{:}1$       & Y / Y          & $N(0,1)$ \\
7 & $H_1$ & $1,2$       & $0,0,0$ & $2$ & $2,0$ & $0$               & $100,100,100$ & $2{:}1$; $2{:}1{:}1$; $2{:}1$       & Y / Y          & Mixture$^{a}$ \\
8 & $H_1$ & $1,2$       & $0,0,0$ & $0$ & $0,0$ & $0$               & $200,1800,200$ & $1{:}1$; $1{:}1{:}98$; $1{:}1$     & Y / N          & $N(0,1)$ \\
9 & $H_1$ & $1,2$       & $0,0,0$ & $0$ & $0,0$ & $0$               & $400,200,200$ & $19{:}1$; $1{:}1{:}1$; $1{:}1$     & N / Y          & $N(0,1)$ \\
\bottomrule
\end{tabular}%
}
\vspace{0.5em}

\footnotesize
$^{a}$ In Scenario 7, $X$ is drawn from a two-component mixture: $0.8\,N(0,0.5^2)+0.2\,N(3,0.5^2)$ in period~1, $0.4\,N(0,0.5^2)+0.6\,N(3,0.5^2)$ in period~2, and $0.2\,N(0,0.5^2)+0.8\,N(3,0.5^2)$ in period~3, giving $E(X\mid S{=}1)=0.6$, $E(X\mid S{=}2)=1.8$, $E(X\mid S{=}3)=2.4$.\\
$^{b}$ The within-period control-to-arm ratio $n_{a,s}/m_{a,s}$ is constant ($1/3$) across periods for both contrasts in this scenario; what varies is the amount of $N_s$ each arm receives, since arm~2 is present only in period~2. This scenario therefore illustrates the $N_s$-vs-$m_{a,s}$.
\end{table}

\subsection{Simulation results}
In Scenarios 1-5 no treatment-period interaction is present, and no covariate-treatment interaction with a covariate distribution unequal across periods. The period-specific effects $\Delta_{a,s}$ are therefore equal across $s \in C_a$, so the ECE, ACA and LACA estimands coincide. Moreover, the working outcome model used in the adjusted estimators (LM, G-computation, AIPW) is correctly specified and every estimator is therefore unbiased for the common target. Figures S1 and S2 (in supplementary material) illustrate the bias for each estimator and scenario. For these adjusted estimators, reweighting period-specific effects by $w_{a,s}^{(P)}$  leaves the ECE, ACA and LACA point estimates unchanged. 
However,  the analysis dataset $\mathcal{I}_D$ does play a role in the adjusted estimators: as anticipated analytically in Section~\ref{sec:exact-aipw}, fitting the outcome model on a larger set estimates the covariate slope and the period effects more precisely and so reduces the variance of the estimator, while restricting it to a single period (LACA fit) inflates the variance. The gain from non-concurrent borrowing is visible but small under a correctly specified model (Scenario 1, arm 2: G-computation on the NCC set has variance $0.022$ versus $0.024$ on ACA), and the loss from the LACA fit is substantial (variance $0.040$ versus $0.024$). See Table S1 in the supplementary material for all results. 

\begin{figure}[h!]
    \centering
    \includegraphics[width=0.7\linewidth]{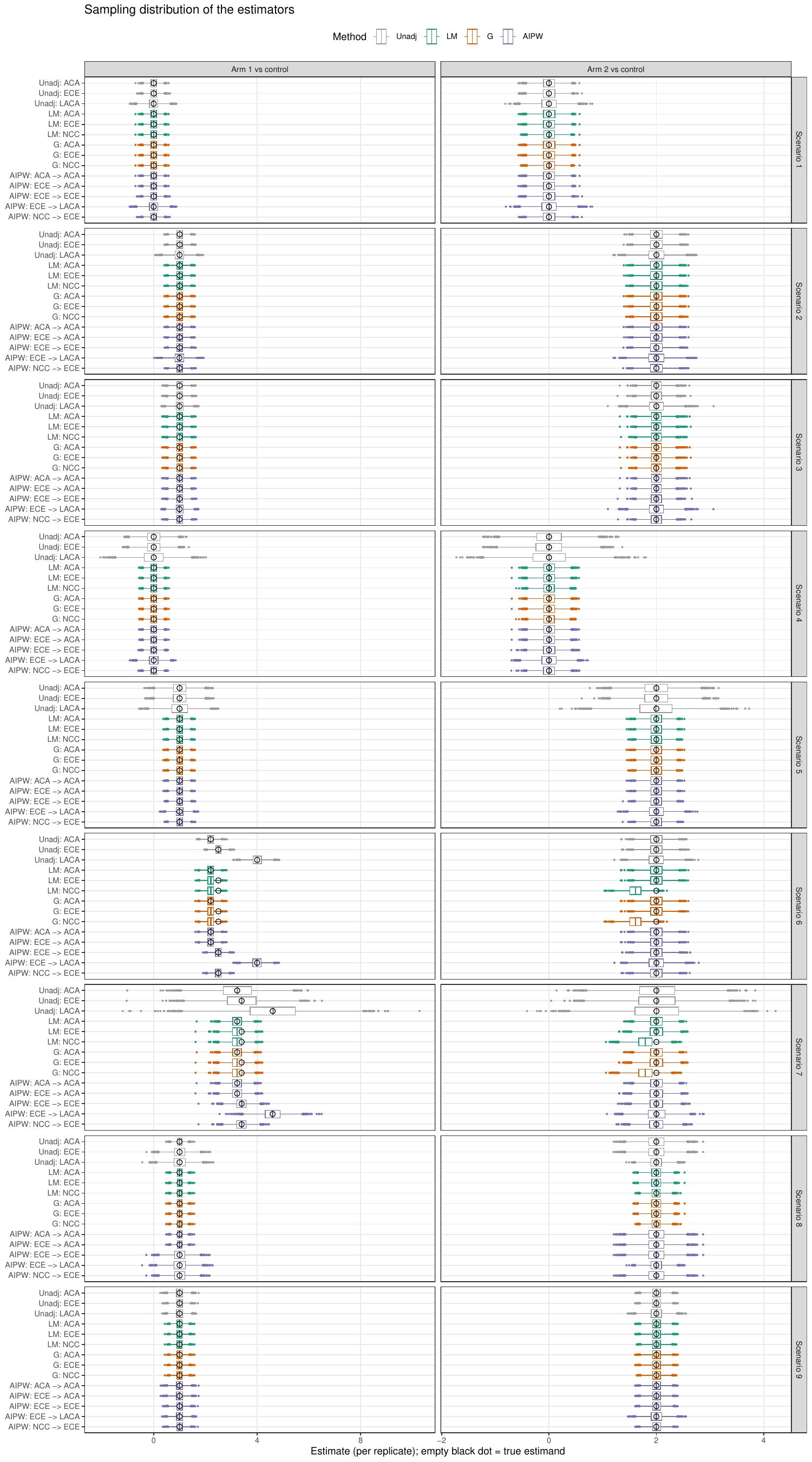}
    \caption{Sampling distribution of the estimators according to the estimation approach and population.}
    \label{fig:fig4_distributions}
\end{figure}

For the unadjusted estimator, while it also remains unbiased, it is observed that in some scenarios, the variance of the estimator is larger compared to the adjusted estimators. This occurs because it does not adjust for covariates. Scenarios 4 and 5 quantify the benefit of covariate adjustment: with a strong covariate ($\beta = 2$) the unadjusted estimator has variance $\approx 0.10$--$0.14$, whereas the adjusted estimators retain variance	$\approx 0.020$--$0.028$. 
Figure \ref{fig:fig4_distributions} presents boxplots of the distribution of the estimators.  The LACA estimator is the least precise, since it discards data from every period but the last. 
Furthermore, simulation results align with the analytic results for the unadjusted estimator in Section \ref{sec:condvar}. Specifically, Scenario 3 confirms the analytical results of ECE-versus-ACA: with $N=(50,150,100)$ the arm 1 comparison has $R_N = 1/3$ and $R_m = 1/2$, and the variance ratio \eqref{eq:var-ratio-ece-aca} equals $1.03$,   closely matching the simulated standard deviations ($0.167$ for ECE versus $0.164$ for ACA).  

Next, we consider Scenario 6, where a treatment-period interaction is present. When $\phi_{1,s} \neq 0$,	the period-specific effects $\Delta_{1,s}$ differ across $s \in C_1$, so the ECE, ACA and LACA estimands no longer coincide as explained in Section \ref{sec:estimands_coincide}. For arm 1, the true values are $2.2$ (ACA), $2.5$ (ECE) and $4.0$ (LACA). 
Figure \ref{fig:fig4_distributions} shows the sampling distribution of the estimators using a boxplot, with the target estimand value highlighted by a black dot to make bias easier to identify. 
The unadjusted estimator, which reweights independent per-period contrasts, recovers each of them ($2.202$, $2.498$, $3.999$). The no-interaction LM and G-computation models are misspecified: they collapse the period-specific effects into a single information-weighted contrast and return $\approx 2.20$ regardless of the target population, so they are biased for the ECE and LACA estimands. They are consistent for ACA, however, as the ratio $n_{a,s}/m_{a,s}$ is constant across periods in $C_a$. AIPW, by contrast, recovers the target estimand (ECE $\approx 2.497$): with randomization probabilities known by design, double robustness rescues the target even under a misspecified outcome model. The same mechanism explains the biased LM/G estimate of the arm 2 effect when	fitted on the non-concurrent set ($1.611$ versus the true $2.0$): borrowing across periods in which arm 1's unmodeled interaction is active biases the period-effect estimates and hence the arm 2 contrast. AIPW on the NCC set is again unbiased ($1.999$).

In Scenario 7, we have a treatment-covariate interaction with covariate shift. Similarly to  Scenario 6, estimands do not coincide because $\psi_1 \neq 0$ and the covariate mean $E(X\mid S=s)$ shifts across periods (see Section \ref{sec:estimands_coincide}). The arm 1 estimands are again distinct (ACA	$3.23$, ECE $3.40$, LACA $4.60$). The additive LM and G-computation models are consistent only for the ACA estimand -- because the allocation ratio $n_{1,s}/m_{1,s}$ is held constant across periods-- and are biased for ECE and LACA (both return $\approx 3.22$). AIPW recovers all three nominal estimands ($3.22$, $3.39$, $4.60$). For arm 2, which has no covariate interaction ($\psi_2 = 0$), LM/G fitted on the NCC set are biased ($1.796$ versus the true $2.0$), whereas AIPW remains valid ($2.000$).  

An important point to note in Scenarios 7-8 is that model misspecification affects not only the point estimator of the treatment effect but also the variance estimators. Figures S3 and S4 in the supplementary material
compare the empirical (Monte Carlo) variance with the mean model-based variance. In the mentioned scenarios, the model-based variance systematically overestimates the true sampling variance.  This is because the misspecification inflates the estimated residual variance without inflating the estimator's actual variance.

The variance formulas behind LM, G-computation, and AIPW all assume the working model $g(E[Y \mid A,X,S])$ captures the conditional mean, so that the residuals $Y_i-\hat p_i^{A_i}$ are pure noise. 
When an interaction is omitted, part of it is absorbed into the residuals as a systematic component that is fixed given arm and period. Because it is fixed given arm and period, it adds no sampling variability to the period-weighted estimator, but the variance formula cannot distinguish a bias-correcting systematic residual from random noise and counts the whole inflated residual as variance.

In Scenario 6, there is a treatment-period interaction for arm 1. The true arm-1 effect is 1 in period 1 and 4 in period 2, but the  model forces a pooled effect coefficient.  
Similarly, in Scenario 7, there is a treatment-covariate interaction with a covariate distribution that shifts across periods. The model omits the interaction term, so arm-1 residuals carry the effect modification. This inflates the estimated residual variance. 
The two misspecifications differ in their reach, which explains which estimators are affected. The treatment-period interaction in Scenario 6 is a between-period phenomenon: misfit appears only when the analysis pools periods with different arm-1 effects. Estimators that never pool across such periods are therefore not affected, meaning: the unadjusted estimator (per-period differences in means, with variance computed within period) and LACA (which uses only the last period, where the effect is homogeneous) both remain calibrated (variance ratios $\approx 1$). Datasets that borrow across periods are instead affected: this is why the arm-2 estimate on the non-concurrent (NCC) dataset is both biased in its point estimate and inflated in its variance. The treatment-covariate interaction in Scenario 7 is a within-period phenomenon: the misfit is present inside every period, so restricting to a single period no longer helps and LACA is now also affected (ratio $\approx 1.5$). 
On the other hand, AIPW is doubly robust for the point estimate  (ECE returns the correct 2.50 in Scenario 6) but double robustness does not extend to the variance estimate (0.0396 vs true 0.0254, ratio 1.56), a phenomenon also discussed by \cite{van2026automated}.  
Finally, the amount of overestimation depends on the variance formula. The classical OLS variance used for both LM and G-computation is inflated whenever the working mean model is misspecified. An alternative would be to use robust sandwich variance, which could be less conservative, but would not completely eliminate overestimation. 

Finally, we discuss Scenario 8 and  Scenario 9, both settings with extreme allocation. In Scenario 8 a dominant competing arm leaves only $m_{1,2} \approx 36$ concurrent patients in a period with $N_2 = 1800$. The arm 1 estimands still coincide (no interaction), but for the unadjusted and AIPW estimators the ECE estimand, whose weights are proportional to $N_s$, places almost all weight on this information-poor period and has variance roughly five times that of ACA ($0.09$ versus $0.017$). The LM and G-computation estimators are not affected: with no interaction they collapse to the weighted contrast and attain variance $0.017$ irrespective of the target population. The same period creates a near-positivity problem for the arm 2 versus control comparison ($18$ controls against $1764$ patients on arm~2): AIPW, which inverts the propensity score, becomes unstable (variance $0.045$ on ACA), while G-computation, which relies only on the outcome model, stays precise (variance $0.015$); here the LACA analysis, which uses only the balanced last period, is actually more precise than ACA for AIPW	($0.020$ versus $0.045$). 

In Scenario 9, the allocation ratio varies across periods, so the ACA weights $w_{a,s}^{\mathrm{ACA}} \propto m_{a,s}$ are no longer inverse-variance optimal: for arm 1 the unadjusted and AIPW estimators have larger variance under ACA than under ECE ($0.031$ versus $0.027$), and the LM/G estimators, which implicitly apply inverse-variance weights, achieve the smallest variance of all ($0.019$). Hence, Scenarios~8 and~9 show that the double-robustness of AIPW comes with fragility under near-positivity, that fixed-weight estimands (ECE, ACA) can be inefficient when the design is unbalanced, and that the model-based estimators --when no interaction is present-- automatically recover an	efficient combination of periods.

\section{Analytic and numerical results for binary endpoints}\label{sec:binary}

The conditions for the concurrent estimands to coincide (Section~\ref{sec:condvar}) were derived for a continuous outcome and a collapsible (linear) effect measure. However, they do not carry over to a binary outcome using logistic regression. 

\subsection{Data-generating model}

For a binary outcome $Y_i \in \{0,1\}$, we replace the identity link of the continuous model \eqref{datagen-model} by the logit link. The analogue of
\eqref{datagen-model} is
\begin{equation}\label{eq:datagen-bin}
	\textrm{logit}\left(  E(Y_i(a) \mid X_i, S_i = s) \right) 
	= \alpha_s + \theta_a + \phi_{a,s} + (\beta + \kappa_s + \psi_a)\, X_i, \qquad X_i \mid S_i = s \sim F_s ,
\end{equation}
where, as before, $\alpha_s$ is the period effect, 
$\theta_a$ is the treatment main
effect, $\phi_{a,s}$ the treatment-period interaction, $\psi_a$ the treatment-covariate interaction, and $F_s$ the period-specific covariate distribution. 

\subsection{Estimands}

Let $\mu_{a,s}$ be the period-specific marginal risk under arm $a$, given by
\begin{equation}\label{eq:marg-risk-bin}
	\mu_{a,s}
	= E_{X \sim F_s}\!\left[ \mathrm{expit}\!\left(\alpha_s + \theta_a + \phi_{a,s} + (\beta+\psi_a)X\right) \right].
\end{equation}
and let $w_{a,s}^{(P)}$ denote the ECE, ACA or LACA population weights as in
\eqref{estimandpop}. In the binary case, we can consider three effect measures of interest.

\textbf{Conditional log-odds ratio}.  
The conditional period-specific effect is the log-odds ratio
\begin{equation}\label{eq:cond-bin}
	\Delta_{a,s}^{\mathrm{cond}} = \theta_a + \phi_{a,s},
\end{equation}
which is the coefficient $\beta_a$ of the logistic model of Section~\ref{sec:reg}. It does not depend on $\alpha_s$ or on $F_s$. In particular, when $\phi_{a,s}=0$ the conditional estimand is constant across periods.

\textbf{Marginal risk difference}. 
By considering period-specific risk differences, the marginal risk-difference estimand keeps the weighted-average form of Section~\ref{sec:estimands} and thus is given by   
\begin{equation}\label{eq:marg-pop-bin}
	\Delta_a^{(P),\mathrm{RD}} = \sum_{s \in C_a} w_{a,s}^{(P)}\, \bigl(\mu_{a,s}-\mu_{0,s}\bigr).
\end{equation}
where the period-specific effect is $\mu_{a,s}-\mu_{0,s}$ and where the superindex RD indicates the scale. 
The risk difference is collapsible, so \eqref{eq:marg-pop-bin} is simultaneously a weighted average of period-specific effects and the contrast of the population-averaged risks $\mu_a^{(P)} = \sum_{s \in C_a} w_{a,s}^{(P)} \mu_{a,s}$.

\textbf{Marginal log-odds ratio}. Given that the  odds ratio is non-collapsible, we can consider two options: the pooled marginal log-odds ratio and the period-averaged marginal log-odds ratio.

The  population-level pooled marginal log-odds ratio is
\begin{equation}\label{eq:mor-pooled}
	\Delta_a^{(P),\mathrm{mOR}}
	= \mathrm{logit}\bigl(\mu_a^{(P)}\bigr) - \mathrm{logit}\bigl(\mu_0^{(P)}\bigr),
	\qquad
	\mu_a^{(P)} = \sum_{s \in C_a} w_{a,s}^{(P)}\, \mu_{a,s},
\end{equation}
that is, the log-odds ratio computed using the risks averaged over the target population. 

The period-averaged marginal log-odds ratio instead applies the form \eqref{estimandpop} directly on the log-odds scale using the period-specific risks,
\begin{equation}\label{eq:mor-avg}
	\widetilde\Delta_a^{(P),\mathrm{mOR}}
	= \sum_{s \in C_a} w_{a,s}^{(P)}\,
	\bigl\{ \mathrm{logit}(\mu_{a,s}) - \mathrm{logit}(\mu_{0,s}) \bigr\}.
\end{equation}

The two coincide when the period-specific marginal risks $\mu_{a,s}$ and $\mu_{0,s}$ are constant over $s \in C_a$, and in particular whenever a single period contributes, that is, for the LACA estimand.


\subsection{When do the concurrent estimands coincide?}\label{sec:bin_estimandscoincide}

The conditions derived in Section~\ref{sec:condvar} for the concurrent estimands ECE, ACA and LACA to coincide were obtained for a continuous outcome and a collapsible (linear) effect measure.  In the linear model, absent a treatment-period interaction and absent a treatment-covariate interaction acting on a shifting covariate mix, $\Delta_{a,s}$ is constant across $s$ and all concurrent estimands are equal. 
However, under \eqref{eq:datagen-bin} this fails. Due to non-linearity, the marginal risk $\mu_{a,s}$ in \eqref{eq:marg-risk-bin} depends on the baseline risk $\alpha_s$ and on the covariate distribution $F_s$. Hence, the marginal period-specific effect $\Delta_{a,s}$ --on either the risk difference or the log-odds ratio scales, and in contrast to the conditional log-odds ratio $\Delta_{a,s}^{\mathrm{cond}}$, which is unaffected by $\alpha_s$ and $F_s$--  varies across periods whenever
\begin{enumerate}
	\item the baseline risk shifts over time ($\alpha_s$ not constant), or
	\item the covariate distribution $F_s$ changes over periods,
\end{enumerate}
even if the conditional log-odds ratio is constant ($\phi_{a,s}=0$) and $\psi_a$ is constant.
Note that the pooled marginal log-odds ratio \eqref{eq:mor-pooled} and the period-averaged marginal log-odds ratio \eqref{eq:mor-avg} differ exactly under the same conditions.  

Hence the ECE, ACA and LACA marginal estimands can differ in the binary case under conditions that leave them equal in the linear case, and the choice of target population is crucial even in the absence of interactions.

\subsection{Estimators}

The estimators of Section~\ref{sec:reg}--\ref{sec:aipw} carry over with the logit link. The logistic regression coefficient estimates the conditional log-odds ratio \eqref{eq:cond-bin}; standardization (G-computation) and AIPW, built from the same logistic outcome model, estimate the marginal effect \eqref{eq:marg-pop-bin} in the target population. As in the continuous case, AIPW remains doubly robust: because the randomization probabilities are known within each period, it is consistent for the nominal marginal estimand even when the logistic outcome model is misspecified.

An important consequence concerns the model-based estimators. In the linear case, when the outcome model contains no treatment-period interaction, the predicted within-period contrasts $\hat p_i^{a}-\hat p_i^{0}$ are identical across periods, so reweighting them by $w_{a,s}^{(P)}$ leaves the estimator unchanged and ECE/ACA/LACA collapse to a single estimate (Section~\ref{sec:condvar}). However, in the binary case, the period-specific standardized contrasts $\hat\Delta_{a,s}^{\mathrm{Stand}}$ in \eqref{eq:stand-s} differ across periods, so reweighting the same fitted model to different target populations produces different estimates. The target population therefore re-enters the analysis through the weights even for G-computation and AIPW, and even without any treatment-period interaction.

\subsection{Numerical illustration}\label{sec:binary-results}

We use the running example of Figure \ref{fig:example} with three periods and two experimental arms and a shared control. We consider three scenarios, all with  conditional log-odds ratios of $\theta_1=0.8$, $\theta_2=1.2$ and no treatment-period interaction ($\phi_{a,s}=0$). Also, unlike the continuous simulations we do not include a covariate-period interaction ($\kappa_s = 0$). The first two scenarios illustrate, separately, the two situations  where the estimands do not match, while the last combines both. 
\begin{itemize}
	\item \textbf{Scenario A (baseline-risk shift).} A stepwise logit-scale time trend, $\alpha_s=(s-1)\lambda$ with $\lambda=1$, so the control-arm risk increases across periods. Period sizes are  unequal  ($N_1=150$, $N_2=450$, $N_3=300$) and allocation is equal within periods ($1:1$ in periods 1
	and 3, $1:1:1$ in period 2). No covariate is included ($\beta=0$).  
	\item \textbf{Scenario B (covariate-distribution shift).} No time trend ($\lambda=0$), equal period sizes ($N_s=300$) and equal allocation, but the distribution $F_s$ of the covariate (with effect $\beta=1$, no treatment-covariate interaction, $\psi_a=0$) shifts across periods. Similarly as in Scenario 7,  $X$ is drawn from a two-component mixture of $N(0,0.5^2)$ and $N(3,0.5^2)$, with the weight on the high-mean component increasing over periods (80\%/20\%, 40\%/60\%, and 20\%/80\% low/high in periods 1, 2, and 3, respectively). 
    \item \textbf{Scenario C (both mechanisms, strong covariate).} A
	stepwise trend is present ($\lambda=0.5$), period sizes are unequal
	($N_1=150$, $N_2=450$, $N_3=300$), allocation in period~2 is unequal
	($2{:}1{:}1$), and the covariate is strongly prognostic ($\beta=1.5$) with the
	same period-varying mixture distribution $F_s$ as in Scenario~B.  
\end{itemize} 

Table \ref{tab:bin-estimands} reports these true marginal risk-difference estimands and Table \ref{tab:bin-estimands-or} reports these true marginal log-odds ratio estimands and the conditional log-odds ratio estimand. 

First, when the covariate has an effect, non-collapsibility means that the marginal log-odds ratios differ from the conditional log-odds ratio (see Scenarios~B and~C). Whereas in scenario A, in which the covariate has no effect, the period-averaged marginal log-odds ratios reproduce the conditional values exactly ($0.800$ and $1.200$). 
Second, the two constructions of the population-level log-odds ratio do not agree. In Scenario~A the pooled values ($0.767$ for arm~1, $1.174$ for arm~2) fall below the period-averaged ones  ($0.800$ and $1.200$). The same appears in Scenarios~B and~C. The two constructions necessarily coincide for LACA, where a single period contributes. 
Third, on the risk-difference scale, the same mechanism makes the period-specific effects heterogeneous, so the ECE, ACA and LACA estimands separate  even though there is no treatment-period interaction. 
Across all three scenarios LACA is the smallest of the three populations, since the last concurrent period is the one with the highest baseline risk and therefore the smallest risk difference, and it is the only estimand that discards the earlier concurrent periods entirely.

\begin{table}[h!]
	\centering
	\caption{True marginal risk-difference estimands and the conditional log-odds ratio ($\theta_a$), for the three scenarios: (A) baseline-risk shift, (B) covariate-distribution shift, and (C) both together with a strongly prognostic covariate. 
    }
	\label{tab:bin-estimands}
	\begin{tabular}{llccc c}
		\toprule
		& & \multicolumn{3}{c}{Marginal risk difference} & Conditional \\
		\cmidrule(lr){3-5}
		Scenario & Contrast & ECE & ACA & LACA & log-OR \\
		\midrule
		A: baseline-risk shift & Arm 1 vs.\ 0 & $0.143$ & $0.148$ & $0.127$ & $0.80$ \\
		(trend, unequal sizes) & Arm 2 vs.\ 0 & $0.134$ & $0.125$ & $0.080$ & $1.20$ \\
		\addlinespace
		B: covariate shift & Arm 1 vs.\ 0 & $0.120$ & $0.126$ & $0.089$ & $0.80$ \\
		(no trend) & Arm 2 vs.\ 0 & $0.102$ & $0.098$ & $0.080$ & $1.20$ \\
		\addlinespace
		C: both, strong covariate & Arm 1 vs.\ 0 & $0.083$ & $0.087$ & $0.064$ & $0.80$ \\
		(trend, unequal alloc) & Arm 2 vs.\ 0 & $0.068$ & $0.065$ & $0.037$ & $1.20$ \\
		\bottomrule
	\end{tabular}
\end{table}

\begin{table}[h!]
	\centering
	\scriptsize
	\caption{True marginal log-odds ratio estimands according to ECE ACA and LACA populations using pooled and period-averaged log-odds ratio, and the conditional log-odds ratio estimand for the three scenarios: (A) baseline-risk shift, (B) covariate-distribution shift, and (C) both together with a strongly prognostic covariate. }
	\label{tab:bin-estimands-or}
	\begin{tabular}{llccc ccc c}
		\toprule
		& & \multicolumn{3}{c}{Marginal log-OR (pooled)} & \multicolumn{3}{c}{Marginal log-OR (period-averaged)} & Conditional \\
		\cmidrule(lr){3-5} \cmidrule(lr){6-8}
		Scenario & Contrast & ECE & ACA & LACA & ECE & ACA & LACA & log-OR \\
		\midrule
		A: baseline-risk shift & Arm 1 vs.\ 0 & $0.767$ & $0.763$ & $0.800$ & $0.800$ & $0.800$ & $0.800$ & $0.80$ \\
		(trend, unequal sizes) & Arm 2 vs.\ 0 & $1.174$ & $1.171$ & $1.200$ & $1.200$ & $1.200$ & $1.200$ & $1.20$ \\
		\addlinespace
		B: covariate shift & Arm 1 vs.\ 0 & $0.629$ & $0.638$ & $0.596$ & $0.639$ & $0.648$ & $0.596$ & $0.80$ \\
		(no trend) & Arm 2 vs.\ 0 & $0.913$ & $0.913$ & $0.922$ & $0.922$ & $0.922$ & $0.922$ & $1.20$ \\
		\addlinespace
		C: both, strong covariate & Arm 1 vs.\ 0 & $0.567$ & $0.568$ & $0.579$ & $0.591$ & $0.594$ & $0.579$ & $0.80$ \\
		(trend, unequal alloc) & Arm 2 vs.\ 0 & $0.895$ & $0.896$ & $0.947$ & $0.919$ & $0.922$ & $0.947$ & $1.20$ \\
		\bottomrule
	\end{tabular}
\end{table}

\section{Discussion}
In this work, we discuss different choices for the target population attribute of estimands in platform trials and assess corresponding estimators. The precise definition of a target population is not a trivial task even in conventional randomized controlled trials (RCTs) \citep{ich2019}. Inclusion and exclusion criteria determine which patients are eligible for a trial, but they do not determine the distribution of patient characteristics within the eligible population. Superpopulation concepts are often used to represent a distribution of patients compatible with the eligibility criteria, for example patients seeking treatment or diagnosis within a specified region and time period. However, recruitment into an RCT is typically not a random sampling process. Trial participants may therefore differ systematically from a broader population of eligible patients, and trial results should be interpreted with this distinction in mind. 

In platform trials, the notion of such a superpopulation is complicated by calendar time. A target population defined over an enrollment interval combines patients arising from a sequence of potentially changing time-specific populations. If there are changes in patient characteristics, disease epidemiology, or standards of care, 
the target population may therefore be better interpreted as a mixture of time-specific populations than as a single static population. Calendar time at diagnosis and treatment is a particular patient characteristic. If calendar time modifies the treatment effect, an average treatment effect over the observed trial period does not automatically predict the treatment effect in future patients. In such settings, the trial results are still informative about treatment effects during the time periods in which the trial was recruiting and, potentially, about their trends over time. However, extrapolation to future patients requires strong assumptions. Therefore, calendar time is different from static baseline effect modifiers because future values of calendar time lie outside the observed trial period.

 Adjustment for calendar time also raises the question of which treatment effect should be targeted. Analyses that condition on calendar period yield period-specific treatment effects, whereas a marginal estimand averages treatment effects over calendar time rather than conditioning on a specific period. Defining such a marginal estimand requires specifying how treatment effects from different periods should be combined and, consequently, which target population the resulting average should represent.

Platform trials make the role of calendar time particularly transparent. The patient population contributing to a specific treatment comparison represents only a subset of the total platform trial population, and the relative weights assigned to different calendar periods by ECE and ACA can depend on the number of treatment arms recruiting in a period and their allocation ratios. In particular, ECE weights periods according to the number of eligible patients enrolled in the platform, whereas ACA weights them according to the number assigned to the treatment under investigation or control. Therefore, the entry or exit of competing treatment arms can lead to differences in the relative contribution of calendar periods between the ACA and ECE populations even if the treatment-control allocation ratio remains constant. Thus, if ACA and ECE estimands apply different period weights, the two estimands can differ due to changes in the marginal treatment effects over time. For marginal treatment effects such changes can occur, even if the treatment effects conditional on observed baseline characteristics are constant over calendar time, but the distribution of effect-modifying baseline characteristics changes over calendar time.

Another consequence is that changes to the platform trial that affect the trial duration may also have an impact on the estimand. For example, if treatment arms are added without a corresponding increase in the overall recruitment rate, the recruitment period may need to be extended to obtain the required sample sizes. If the patient characteristics or marginal treatment effects change over calendar time, this can change in the population over which the treatment effect is averaged and therefore may change the estimand. Similarly, changes in recruitment speed will change the proportions of patients recruited at different calendar times and can therefore have an impact on the estimand if time trends are present.

It is important to distinguish the target population and the data used for estimation. For example, one may target a concurrent estimand such as ECE or ACA but use non-concurrent observations to reduce the standard error of treatment effect estimators by improving the estimation of the control response or the estimation of prognostic covariate effects. This does not change the target population. 

For continuous outcomes with a linear effect measure, ECE, ACA, and LACA coincide under the model considered in Section \ref{sec:rescont} if 
the treatment-period interaction is constant across periods and either there is no treatment-by-covariate interaction or the relevant covariate distribution does not change over time. If ECE, ACA, and LACA coincide, the choice of  estimators can be based on precision and robustness. For unadjusted estimators, the weighting of the ACA estimator is variance-optimal under the assumption of constant allocation ratio and constant residual variance. 
For the adjusted estimators in the absence of treatment-period and treatment-covariate interactions, the target population leave the point estimate unchanged, and a larger analysis set  (ECE or NCC rather than ACA) can reduce the standard error of treatment effect estimators by improving estimation of the control response and covariate effects. When a treatment-period interaction is present or a treatment–covariate interaction combined with a covariate distribution that shifts across periods, the working model is misspecified, and the linear regression and G-computation estimators are biased when using non-concurrent data, whereas AIPW recovers the intended estimand for every analysis set because the randomization probabilities are known by design. Misspecification also affects the variance estimators, whose average exceeds the empirical variance of the estimates, so that inference becomes conservative.

Under the assumption that the marginal treatment effect varies over calendar time, the choice among ECE, ACA, and LACA will depend on the objective of the trial. If the main objective is to inform treatment decisions for patients in the near future, an estimand referring to the most recent patient population, such as LACA, may be particularly relevant. 

In nonlinear models, such as those used for binary outcomes, an additional issue arises. Even if the conditional treatment effect is constant over time, the marginal treatment effect can vary across periods when the baseline response or the covariate distribution changes. For example, under a logistic outcome model with a constant conditional log-odds ratio, both marginal risk differences and marginal odds ratios can change with the baseline risk or the covariate distribution. Thus, differences between ECE, ACA, and LACA in this setting can occur even if no treatment-by-period interaction exists. This is important for the choice of estimand because the different populations then correspond to genuinely different treatment effects even under a constant conditional treatment effect. 

An issue not considered in this manuscript is the setting in which inclusion and exclusion criteria differ across treatments. In this case, treatment-specific eligibility may lead to different target populations for different comparisons. In addition, the introduction of a new treatment arm may change the distribution of patient characteristics among participants available for other comparisons because some patients may be eligible for a larger set of treatment arms than others. The extent of such changes depends on the recruitment and randomization procedures in the platform trial. The same phenomenon can occur when independent trials with overlapping eligibility criteria recruit concurrently at the same centers, but it is especially visible in platform trials. Methods addressing such population shifts have been proposed \cite{degtiar2023review}, and the extension of these approaches in the presence of calendar time trends is left for future work.

\section*{Funding}

M. Bofill Roig is a Serra Húnter Fellow and was additionally supported by Grant PID2023-148033OB-C21 funded by MICIU/AEI/10.13039/501100011033 and by FEDER/UE.

\section*{Conflict of interest}

The authors declare no potential conflict of interest.

\appendix

\section{Derivation of the distribution of the AIPW} \label{app:der}
We start by writing the conditional model, given the observed covariate values \(x_i\), in compact matrix form. Let
\[
\mathbf Y
=
\begin{pmatrix}
\mathbf Y_{0,1}^{\top}&
\mathbf Y_{1,1}^{\top}&
\mathbf Y_{0,2}^{\top}&
\mathbf Y_{1,2}^{\top}&
\mathbf Y_{2,2}^{\top}&
\mathbf Y_{0,3}^{\top}&
\mathbf Y_{2,3}^{\top}
\end{pmatrix}^{\top},
\]
where \(\mathbf Y_{a,s}\) denotes the vector of observed outcomes in treatment-period cell \((a,s)\). Conditional on the covariates and the
realized treatment-period cell membership, the model can be written as
\[
\mathbf Y=\mathcal X\vartheta+\varepsilon,
\qquad
\varepsilon\mid \mathcal X\sim N(0,\bm\Sigma),
\]
where \(\mathcal X\) is the fixed design matrix induced by the observed treatment-period cells and covariate values, and \(\bm\Sigma\) is the
corresponding block-diagonal covariance matrix with blocks $\sigma_{a,s}^2\cdot I_{n_{a,s}}$.

The AIPW estimator is constructed from two components: observed outcomes entering through the inverse-probability weighted residual correction and fitted potential-outcome predictions obtained from the working outcome model. Since the outcome model is linear and fitted by generalized least squares, the fitted parameter vector \(\hat\vartheta\) is itself a linear function of \(\mathbf Y\). Hence the fitted predictions are also linear functions of \(\mathbf Y\). Consequently, after collecting terms, the population-level AIPW estimator can be written as a single linear contrast
\[
\hat\Delta_a^{(P),\mathrm{AIPW}}
=
\ell_a^{(P)\top}\mathbf Y .
\]
The aim of the following derivation is to make the contrast vector \(\ell_a^{(P)}\) explicit. Once this vector is identified, the exact conditional distribution of the estimator follows immediately from the Gaussian linear model:
\[
\hat\Delta_a^{(P),\mathrm{AIPW}}\mid \mathcal X
\sim
N\left(
\ell_a^{(P)\top}\mathcal X\vartheta,\,
\ell_a^{(P)\top}\bm\Sigma \ell_a^{(P)}
\right).
\]

Consider the following identifying restrictions:
$$\qquad
\theta_0=0,\quad \psi_0=0,\quad \kappa_1=0,\quad
\phi_{0,s}=0\ \text{for all }s,\\
\phi_{1,1}=0,\qquad \phi_{2,2}=0.
$$
Thus, $\alpha_s  = E(Y_i\mid A_i=0,S_i=s,X_i=0)$ and
$$\vartheta^{\top}
= (
\alpha_1 \qquad
\alpha_2 \qquad
\alpha_3 \qquad
\theta_1 \qquad
\theta_2 \qquad
\phi_{1,2} \qquad
\phi_{2,3} \qquad
\beta \qquad
\kappa_2 \qquad
\kappa_3 \qquad
\psi_1 \qquad
\psi_2 ).$$

\[ 
\mathcal{X}
= 
\begin{pmatrix}
\mathbf{1}_{0,1} & 0 & 0 & 0 & 0 & 0 & 0 & x_{0,1} & 0 & 0 & 0 & 0 \\
\mathbf{1}_{1,1} & 0 & 0 & \mathbf{1}_{1,1} & 0 & 0 & 0 & x_{1,1} & 0 & 0 & x_{1,1} & 0 \\
0 & \mathbf{1}_{0,2} & 0 & 0 & 0 & 0 & 0 & x_{0,2} & x_{0,2} & 0 & 0 & 0 \\
0 & \mathbf{1}_{1,2} & 0 & \mathbf{1}_{1,2} & 0 & \mathbf{1}_{1,2} & 0 & x_{1,2} & x_{1,2} & 0 & x_{1,2} & 0 \\
0 & \mathbf{1}_{2,2} & 0 & 0 & \mathbf{1}_{2,2} & 0 & 0 & x_{2,2} & x_{2,2} & 0 & 0 & x_{2,2} \\
0 & 0 & \mathbf{1}_{0,3} & 0 & 0 & 0 & 0 & x_{0,3} & 0 & x_{0,3} & 0 & 0 \\
0 & 0 & \mathbf{1}_{2,3} & 0 & \mathbf{1}_{2,3} & 0 & \mathbf{1}_{2,3} & x_{2,3} & 0 & x_{2,3} & 0 & x_{2,3}
\end{pmatrix}. 
\]

\[ 
Y\mid \mathcal X
\sim
N(\mathcal X\vartheta,\Sigma).
\]

\[
V_{\vartheta}
=
(\mathcal X^{\top}\Sigma^{-1}\mathcal X)^{-1},
\qquad
G
=
V_{\vartheta}\mathcal X^{\top}\Sigma^{-1}.
\]

\[
\begin{pmatrix}
Y\\
\hat\vartheta
\end{pmatrix}
\Bigm|\mathcal X
\sim
N
\left[
\begin{pmatrix}
\mathcal X\vartheta\\
\vartheta
\end{pmatrix},
\begin{pmatrix}
\Sigma & \mathcal X V_{\vartheta}\\
V_{\vartheta}\mathcal X^{\top} & V_{\vartheta}
\end{pmatrix}
\right].
\]

Now define the model row for a hypothetical treatment $b$ in period $s$ with covariate value $x_i$ by
$$h(b,s,x_i)^{\top}\vartheta
=
\hat m_b(s,x_i).$$
For example,

\[
\begin{aligned}
h(0,1,x_i)^{\top}
&=
(1,0,0,0,0,0,0,x_i,0,0,0,0),\\
h(1,1,x_i)^{\top}
&=
(1,0,0,1,0,0,0,x_i,0,0,x_i,0),\\
h(0,2,x_i)^{\top}
&=
(0,1,0,0,0,0,0,x_i,x_i,0,0,0),\\
h(1,2,x_i)^{\top}
&=
(0,1,0,1,0,1,0,x_i,x_i,0,x_i,0),\\
h(2,2,x_i)^{\top}
&=
(0,1,0,0,1,0,0,x_i,x_i,0,0,x_i),\\
h(0,3,x_i)^{\top}
&=
(0,0,1,0,0,0,0,x_i,0,x_i,0,0),\\
h(2,3,x_i)^{\top}
&=
(0,0,1,0,1,0,1,x_i,0,x_i,0,x_i).
\end{aligned}
\]

Let $R_P$ be the selection matrix extracting the observations in the target population $\mathcal I_{(P)}$, so that
$$Y_P=R_PY.$$
For treatment $a$, define $H_a^{(P)}$ and $H_0^{(P)}$ as the matrices with rows
$$h(a,S_i,x_i)^{\top}
\quad\text{and}\quad
h(0,S_i,x_i)^{\top},
\qquad i\in\mathcal I_{(P)}.$$
Thus
\[ 
\hat p^a
=
H_a^{(P)}\hat\vartheta,
\qquad
\hat p^0
=
H_0^{(P)}\hat\vartheta.
\]

Define the diagonal inverse-probability matrices

\[
D_a^{(P)}
=
\operatorname{diag}
\left\{
\frac{I(A_i=a)}{\pi_{a,S_i}^{(P)}}:
i\in\mathcal I_{(P)}
\right\},
\qquad
D_0^{(P)}
=
\operatorname{diag}
\left\{
\frac{I(A_i=0)}{\pi_{0,S_i}^{(P)}}:
i\in\mathcal I_{(P)}
\right\}.
\]

Here, for example,
$$\pi_{b,s}^{(P)}=\frac{n_{b,s}^{(P)}}{n_{\cdot,s}^{(P)}},$$
or the corresponding rescaled probability in the ACA population.

The vector of AIPW pseudo-outcomes is

\[
\hat\phi_a
=
D_a^{(P)}(Y_P-H_a^{(P)}\hat\vartheta)
-
D_0^{(P)}(Y_P-H_0^{(P)}\hat\vartheta)
+
(H_a^{(P)}-H_0^{(P)})\hat\vartheta.
\] 

Substituting \(\hat\vartheta=G\mathbf Y\), this becomes
\[
\hat{\bm\phi}_a
=
M_a^{(P)}\mathbf Y,
\]
where
\[
M_a^{(P)}
=
\left(D_a^{(P)}-D_0^{(P)}\right)R_P
+
\left\{
\left(I-D_a^{(P)}\right)H_a^{(P)}
-
\left(I-D_0^{(P)}\right)H_0^{(P)}
\right\}G .
\]

Let 
\[
q_{a,i}^{(P)}
=
\frac{I(S_i\in C_a)}{n_{C_a}^{(P)}},
\qquad
i\in\mathcal I_{(P)}.
\] be the "weight" of individual $i$ in the calculation of the AIPW estimate $\hat{\Delta}_{a,s}^{(P)}$ and $\mathbf{q}_a^{(P)}$ be the corresponding vector. 

Then
\[
\hat\Delta_a^{(P),\mathrm{AIPW}}
=
\mathbf{q}_{a}^{(P)\top}\hat{\bm{\phi}}_a
=
\mathbf{q}_{a}^{(P)\top}M_a^{(P)}\mathbf Y.
\]

Hence the AIPW estimator is the linear contrast
\[
\hat\Delta_a^{(P),\mathrm{AIPW}}
=
\ell_a^{(P)\top}\mathbf Y,
\qquad
\ell_a^{(P)}
=
M_a^{(P)\top}q_a^{(P)}.
\]

Therefore, conditional on the fixed design \(\mathcal X\),
\[
\hat\Delta_a^{(P),\mathrm{AIPW}}
\mid \mathcal X
\sim
N\left(
\ell_a^{(P)\top}\mathcal X\vartheta,
\,
\ell_a^{(P)\top}\bm\Sigma\ell_a^{(P)}
\right).
\]

\normalsize

\bibliography{refs}

@article{degtiar2023review,
  title={A review of generalizability and transportability},
  author={Degtiar, Irina and Rose, Sherri},
  journal={Annual Review of Statistics and Its Application},
  volume={10},
  number={1},
  pages={501--524},
  year={2023},
  publisher={Annual Reviews}
}

@article{robins1994estimation,
  title={Estimation of regression coefficients when some regressors are not always observed},
  author={Robins, James M and Rotnitzky, Andrea and Zhao, Lue Ping},
  journal={Journal of the American statistical Association},
  volume={89},
  number={427},
  pages={846--866},
  year={1994},
  publisher={Taylor \& Francis}
}

@book{ich2019,
   author = {ICH},
   title = {{International Council for Harmonisation} Topic E9(R1) on ``Estimands and Sensitivity Analysis in Clinical Trials''},
   url = {www.ich.org},
   year = {2019},
   note = {Last checked: 2026-09-11}
}

@article{robins1986new,
  title={A new approach to causal inference in mortality studies with a sustained exposure period—application to control of the healthy worker survivor effect},
  author={Robins, James},
  journal={Mathematical modelling},
  volume={7},
  number={9-12},
  pages={1393--1512},
  year={1986},
  publisher={Elsevier}
}

@article{saville2016efficiencies,
  title={Efficiencies of platform clinical trials: a vision of the future},
  author={Saville, Benjamin R and Berry, Scott M},
  journal={Clinical Trials},
  volume={13},
  number={3},
  pages={358--366},
  year={2016},
  publisher={SAGE Publications Sage UK: London, England}
}

@article{berry2015platform,
  title={The platform trial: an efficient strategy for evaluating multiple treatments},
  author={Berry, Scott M and Connor, Jason T and Lewis, Roger J},
  journal={Jama},
  volume={313},
  number={16},
  pages={1619--1620},
  year={2015}
}

@article{koenig2024current,
  title={Current state-of-the-art and gaps in platform trials: 10 things you should know, insights from EU-PEARL},
  author={Koenig, Franz and Spiertz, C{\'e}cile and Millar, Daniel and Rodr{\'\i}guez-Navarro, Sarai and Mach{\'\i}n, N{\'u}ria and Van Dessel, Ann and Genesc{\`a}, Joan and Peric{\`a}s, Juan M and Posch, Martin and S{\'a}nchez-Montalva, Adrian and others},
  journal={EClinicalMedicine},
  volume={67},
  year={2024},
  publisher={Elsevier}
}

@techreport{fda2023covariate,
  author      = {{U.S. Food and Drug Administration}},
  title       = {Adjusting for Covariates in Randomized Clinical Trials for Drugs and Biological Products: Guidance for Industry},
  institution = {U.S. Department of Health and Human Services, Food and Drug Administration, Center for Drug Evaluation and Research (CDER)},
  year        = {2023},
  month       = {May},
  type        = {Guidance for Industry},
  url         = {https://www.fda.gov/regulatory-information/search-fda-guidance-documents/adjusting-covariates-randomized-clinical-trials-drugs-and-biological-products}
}

@article{van2026automated,
  title={Automated, efficient and model-free inference for randomized clinical trials via data-driven covariate adjustment},
  author={Van Lancker, Kelly and D{\'\i}az, Iv{\'a}n and Vansteelandt, Stijn},
  journal={The American Statistician},
  number={just-accepted},
  pages={1--26},
  year={2026},
  publisher={Taylor \& Francis}
}

@article{van2024covariate,
  title={Covariate adjustment in randomized controlled trials: General concepts and practical considerations},
  author={Van Lancker, Kelly and Bretz, Frank and Dukes, Oliver},
  journal={Clinical Trials},
  volume={21},
  number={4},
  pages={399--411},
  year={2024},
  publisher={SAGE Publications Sage UK: London, England}
}

@article{qian2024estimands,
  title={From Estimands to Robust Inference of Treatment Effects in Platform Trials},
  author={Qian, Yuhan and Yi, Yifan and Shao, Jun and Yi, Yanyao and Levin, Gregory and Mayer-Hamblett, Nicole and Heagerty, Patrick J and Ye, Ting},
  journal={arXiv preprint arXiv:2411.12944},
  year={2024}
}

@article{daniel2021making,
  title={Making apples from oranges: Comparing noncollapsible effect estimators and their standard errors after adjustment for different covariate sets},
  author={Daniel, Rhian and Zhang, Jingjing and Farewell, Daniel},
  journal={Biometrical Journal},
  volume={63},
  number={3},
  pages={528--557},
  year={2021},
  publisher={Wiley Online Library}
}

@article{santacatterina2025identification,
  title={Identification and Estimation of Causal Effects Using Non-Concurrent Controls in Platform Trials},
  author={Santacatterina, Michele and Giron, Federico Macchiavelli and Zhang, Xinyi and D{\'\i}az, Iv{\'a}n},
  journal={Statistics in Medicine},
  volume={44},
  number={6},
  pages={e70017},
  year={2025},
  publisher={Wiley Online Library}
}

@article{zhan2025integrated,
  title={An integrated and coherent framework for point estimation and hypothesis testing with concurrent controls in platform trials},
  author={Zhan, Tianyu and Zhang, Jane and Shu, Lei and Gu, Yihua},
  journal={Statistics in Medicine},
  volume={44},
  number={15-17},
  pages={e70196},
  year={2025},
  publisher={Wiley Online Library}
}

@article{burgwinkel2023use,
  title={On the use of non-concurrent controls in platform trials: a scoping review.},
  author={Burgwinkel, C and Garczarek, U and Koenig, F and Posch, M and Nguyen, Q and Hees, K and others},
  journal={Trials},
  volume={24},
  number={1},
  pages={408--408},
  year={2023}
}

@article{krotka2025statistical,
  title={Statistical Modeling to Adjust for Time Trends in Adaptive Platform Trials Utilizing Non-Concurrent Controls},
  author={Krotka, Pavla and Posch, Martin and Gewily, Mohamed and H{\"o}glinger, G{\"u}nter and Bofill Roig, Marta},
  journal={Biometrical journal},
  volume={67},
  number={3},
  pages={e70059},
  year={2025},
  publisher={Wiley Online Library}
}

@article{marschner2024analysis,
  title={Analysis of nonconcurrent controls in adaptive platform trials: separating randomized and nonrandomized information},
  author={Marschner, Ian C and Schou, I Manjula},
  journal={Biometrical Journal},
  volume={66},
  number={6},
  pages={e202300334},
  year={2024},
  publisher={Wiley Online Library}
}

@article{sridhara2022use,
  title={Use of nonconcurrent common control in master protocols in oncology trials: report of an American statistical association biopharmaceutical section open forum discussion},
  author={Sridhara, Rajeshwari and Marchenko, Olga and Jiang, Qi and Pazdur, Richard and Posch, Martin and Berry, Scott and Theoret, Marc and Shen, Yuan Li and Gwise, Thomas and Hess, Lorenzo and others},
  journal={Statistics in Biopharmaceutical Research},
  volume={14},
  number={3},
  pages={353--357},
  year={2022},
  publisher={Taylor \& Francis}
}

@article{bofill2022model,
  title={On model-based time trend adjustments in platform trials with non-concurrent controls},
  author={Bofill Roig, Marta and Krotka, Pavla and Burman, Carl-Fredrik and Glimm, Ekkehard and Gold, Stefan M and Hees, Katharina and Jacko, Peter and Koenig, Franz and Magirr, Dominic and Mesenbrink, Peter and others},
  journal={BMC medical research methodology},
  volume={22},
  number={1},
  pages={228},
  year={2022},
  publisher={Springer}
}

@article{saville2022bayesian,
  title={The Bayesian time machine: accounting for temporal drift in multi-arm platform trials},
  author={Saville, Benjamin R and Berry, Donald A and Berry, Nicholas S and Viele, Kert and Berry, Scott M},
  journal={Clinical Trials},
  volume={19},
  number={5},
  pages={490--501},
  year={2022},
  publisher={SAGE Publications Sage UK: London, England}
}

@article{kopp2020power,
  title={Power gains by using external information in clinical trials are typically not possible when requiring strict type I error control},
  author={Kopp-Schneider, Annette and Calderazzo, Silvia and Wiesenfarth, Manuel},
  journal={Biometrical Journal},
  volume={62},
  number={2},
  pages={361--374},
  year={2020},
  publisher={Wiley Online Library}
}

@article{tsiatis2008covariate,
  title={Covariate adjustment for two-sample treatment comparisons in randomized clinical trials: a principled yet flexible approach},
  author={Tsiatis, Anastasios A and Davidian, Marie and Zhang, Min and Lu, Xiaomin},
  journal={Statistics in medicine},
  volume={27},
  number={23},
  pages={4658--4677},
  year={2008},
  publisher={Wiley Online Library}
}

@article{guo2024treatment,
  title={Treatment comparisons in adaptive platform trials adjusting for temporal drift},
  author={Guo, Beibei and Wang, Li and Yuan, Ying},
  journal={Statistics in biopharmaceutical research},
  volume={16},
  number={3},
  pages={361--370},
  year={2024},
  publisher={Taylor \& Francis}
}

@article{lee2020including,
  title={Including non-concurrent control patients in the analysis of platform trials: is it worth it?},
  author={Lee, Kim May and Wason, James},
  journal={BMC Medical Research Methodology},
  volume={20},
  number={1},
  pages={165},
  year={2020},
  publisher={Springer}
}

@article{bannick2026covariate,
  title={Covariate adjustment in randomized clinical trials: From general theory to practical insights},
  author={Bannick, Marlena S and Yi, Yanyao and Ye, Ting},
  journal={Clinical Trials},
  volume={23},
  number={3},
  pages={272--281},
  year={2026},
  publisher={SAGE Publications Sage UK: London, England}
}

\newpage

\includepdf[pages=-]{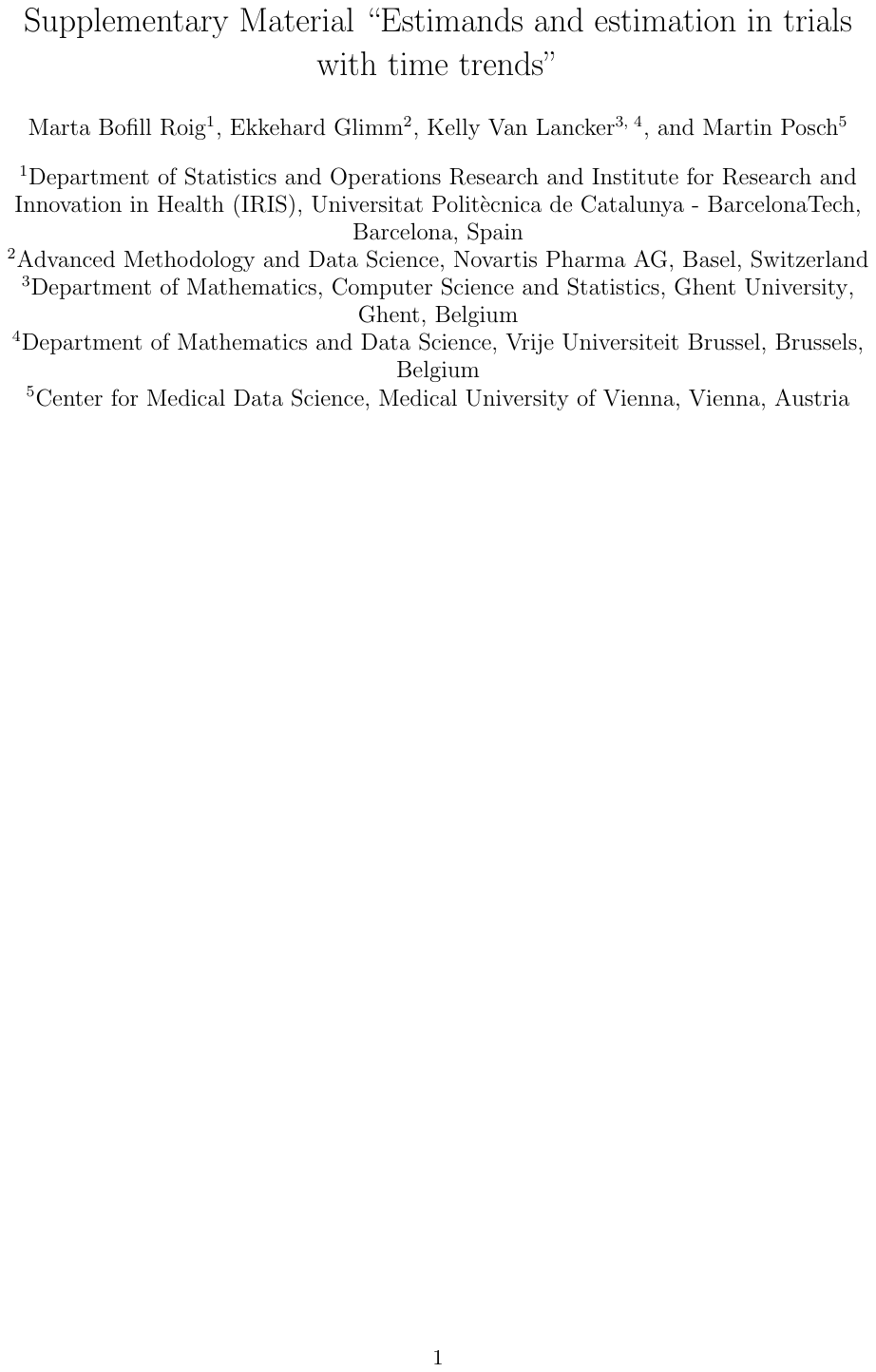}

\clearpage

\end{document}